\documentclass[11pt]{scrartcl}
\makeatletter
\DeclareOldFontCommand{\rm}{\normalfont\rmfamily}{\mathrm}
\DeclareOldFontCommand{\sf}{\normalfont\sffamily}{\mathsf}
\DeclareOldFontCommand{\tt}{\normalfont\ttfamily}{\mathtt}
\DeclareOldFontCommand{\bf}{\normalfont\bfseries}{\mathbf}
\DeclareOldFontCommand{\it}{\normalfont\itshape}{\mathit}
\DeclareOldFontCommand{\sl}{\normalfont\slshape}{\@nomath\sl}
\DeclareOldFontCommand{\sc}{\normalfont\scshape}{\@nomath\sc}
\makeatother
\usepackage{epsf,amsmath,amssymb,graphicx,scalefnt}
\usepackage[bf,format=plain]{caption}
\usepackage{rotating,enumitem,fancyhdr,slashed}
\usepackage{filemod}
\usepackage[nosort,noadjust]{cite}
\usepackage[final]{showlabels}
\usepackage{longtable}
\usepackage[affil-it]{authblk}
\usepackage[usenames,dvipsnames]{color}
\usepackage{hyperref}

\newcommand{\vhnnlo}{\texttt{vh@nnlo}}
\newcommand{\hawk}{\texttt{HAWK}}
\newcommand{\mcfm}{\texttt{MCFM}}
\newcommand{\thdm}{\abbrev{2HDM}}
\newcommand{\vlq}{\abbrev{VLQ}}

\newcommand{\zw}{\ensuremath{Z\!W}}
\newcommand{\zh}{\ensuremath{Z\!H}}
\newcommand{\wh}{\ensuremath{W\!H}}
\newcommand{\vh}{\ensuremath{V\!H}}

\newcommand{\ggzh}{\ensuremath{gg\to \zh}}
\newcommand{\bbzh}{\ensuremath{b{\bar b}\to \zh}}

\newcommand{\pt}{\ensuremath{p_T}}
\newcommand{\pth}{\ensuremath{p_{T}^H}}
\newcommand{\mvh}{\ensuremath{M_{\vh}}}
\newcommand{\mzh}{\ensuremath{M_{\zh}}}
\newcommand{\dy}{\text{\abbrev{DY}}}
\newcommand{\as}{\alpha_s}
\newcounter{notecount}

\newcommand{\RZW}[2]{\ensuremath{R^{\zw#1}_{#2}}}
\newcommand{\RRZW}[2]{\ensuremath{R_{R#2}^{\zw#1}}}
\newcommand{\RZWDY}[2]{\ensuremath{R_\text{\dy#2}^{\zw#1}}}

\newcommand{\higgsstrahlung}{Higgs-Strahlung}
\newcommand{\citere}[1]{Ref.\,\cite{#1}}
\newcommand{\citeres}[1]{Refs.\,\cite{#1}}

\newcommand{\abbrev}[1]{{\scalefont{.9}#1}}

\newcommand{\eqn}[1]{Eq.\,(\ref{#1})}
\newcommand{\noeqn}[1]{(\ref{#1})}
\newcommand{\eqns}[1]{Eqs.\,(\ref{#1})}
\newcommand{\fig}[1]{Fig.\,\ref{#1}}
\newcommand{\figs}[1]{Figs.\,\ref{#1}}

\newcommand{\sct}[1]{Sect.\,\ref{#1}}

\newcommand{\dd}{{\rm d}}

\newcommand{\order}[1]{{\cal O}(#1)}
\newcommand{\lhc}{\abbrev{LHC}}
\newcommand{\qcd}{\abbrev{QCD}}
\newcommand{\sm}{\abbrev{SM}}

\newcommand{\bsm}{\abbrev{BSM}}
\newcommand{\pdf}{\abbrev{PDF}}
\newcommand{\lo}{\abbrev{LO}}
\newcommand{\nlo}{\abbrev{NLO}}
\newcommand{\nnlo}{\abbrev{NNLO}}
\newcommand{\nll}{\abbrev{NLL}}

\newcommand{\muF}{\mu_\text{F}}
\newcommand{\muR}{\mu_\text{R}}
\newcommand{\mhiggs}{M_\text{H}}
\newcommand{\mtop}{M_\text{t}}

\newcommand{\RHheaderline}{\textit{April 2018}\\TTK-17-48\\Nikhef 2018-026}
\fancypagestyle{firstpage}
{
  
  \fancyhead[R]{\RHheaderline}
}
\title{Exploiting the WH/ZH symmetry in the search for New Physics}
\author{R.V. Harlander,$^{1}$ J.~Klappert,$^{1}$ C. Pandini,$^{2}$ and A.~Papaefstathiou$^{3,4}$}
\affil{$^{1}$ TTK, RWTH Aachen University, D-52056 Aachen, Germany.\\
$^{2}$ EP Department, CERN, CH-1211 Geneva 23, Switzerland.\\
$^{3}$ Institute for Theoretical Physics Amsterdam and Delta Institute
  for Theoretical Physics, University of Amsterdam,
  Science Park 904, 1098 XH Amsterdam, The Netherlands.\\
  $^{4}$ Nikhef, Theory Group, Science Park 105,
  1098 XG, Amsterdam, The Netherlands.}
\date{}
\begin{document}
\maketitle
\begin{abstract}
  We suggest to isolate the loop-induced gluon-initiated component (\ggzh) for associated
  $\zh$ production by using the similarity of the Drell-Yan-like
  component for $\zh$ production to the $\wh$ process. We argue that the cross-section
  ratio of the latter two processes can be predicted with high
  theoretical accuracy. Comparing it to the experimental $\zh/\wh$ cross-section ratio
  should allow to probe for New Physics in the \ggzh\ component at the
  \abbrev{HL-LHC}. We consider typical \bsm{} scenarios in order to
  exemplify the effect they would have on the proposed observable.
\end{abstract}
\thispagestyle{firstpage}

%- }}}
%- {{{ body:

%- {{{ Introduction:

\newpage
\section{Introduction}\label{sec:intro}

The Higgs boson provides a new probe for physics beyond the Standard
Model (\sm). A precise measurement of its couplings to the \sm{}
particles is certainly one of the most promising ways to search for
deviations from the \sm{}. The tree-level couplings of the \sm{} Higgs
are determined solely by the particle masses and the vacuum expectation
value $v\approx 246$\,GeV; a global fit to these couplings yields good
agreement with the \sm{} predictions within current experimental
uncertainties, see e.g.\ \citere{Khachatryan:2016vau}.  Couplings to
massless particles like the photon or gluons are necessarily
loop-induced, which allows for New Physics to affect
the numerical value or the Lorentz structure of these couplings in a
significant way.

In fact, the loop-induced couplings to photons as well as to gluons were
essential to the actual discovery of the Higgs boson, for example
through $gg\to H\to \gamma\gamma$\,\cite{Aad:2012tfa,
  Chatrchyan:2012xdj}. The good agreement with the theoretical
prediction of this process within the \sm{} leaves little room for any
large impact of New Physics here (for comprehensive reviews on Higgs
physics at the Large Hadron Collider (\lhc),
see \citeres{Dittmaier:2011ti,Dittmaier:2012vm,%
  Heinemeyer:2013tqa,deFlorian:2016spz}).

Associated $\vh$ production, or \higgsstrahlung\ for short, is one of
the main production modes for Higgs bosons at the \lhc. Despite its
rather small cross section, its feature of providing a tag through the
electro-weak gauge boson in the final state recently allowed the first
observation of the Higgs decay to bottom quarks, which is swamped by
background $b\bar b$ production in other major Higgs production modes. 
Focusing on boosted-Higgs events and advanced jet-substructure analyses 
is a promising direction to further separate the signal from the background\,\cite{Butterworth:2008iy}.

Another unique feature of the \higgsstrahlung\ process $pp\to \vh$ is
its appearance in two variants: $V=W$ and $V=Z$. In the \sm{}, the
amplitudes are related through next-to-leading order (\nlo) \qcd{} by
well established symmetry properties of the \sm{}. At
next-to-next-to-leading order (\nnlo)\ \qcd, however, these two concepts
are no longer sufficient to relate $\wh$ to $\zh$ production. This is
mostly due to a loop-induced contribution to $\zh$ production whose
leading-order (\lo) partonic amplitude is given by \ggzh. The
corresponding Feynman diagrams contain either boxes or triangles of
bottom or top quarks, see \fig{fig:ggzhlo} (lighter-quark contributions
are numerically negligible in general).  In the \sm{}, the box and
triangle contributions interfere destructively, which leads to an
enhanced sensitivity to physics beyond the \sm{} (\bsm).

Quite generally, loop-induced processes are particularly sensitive to
New Physics, since its effects are likely to be of the same order as the
\sm\ process in this case. The sub-process \ggzh, however, is only one
contribution to the general \higgsstrahlung\ process of $\zh$
production, albeit a separately-finite and gauge-independent
one. Moreover, it is suppressed by two powers of the strong coupling
constant $\as$ with respect to the dominant $q\bar q$-initiated contribution.

It would thus be desirable to separate event samples which are due to
the ``Drell-Yan like'' production mechanism, where at \lo{} the Higgs is
radiated off an off-shell $Z$ boson produced in $q\bar q$ annihilation,
from the ones due to the gluon-initiated process \ggzh. First steps in
this direction have been taken in \citeres{Harlander:2013mla} and
\cite{Englert:2013vua}. In the latter paper, it was pointed out that the
relative contribution of the \ggzh\ process to $\zh$ production depends
strongly on the kinematical region of the final state.  For example,
while it constitutes only about 6\% of the total cross section (at order
$\as^2$), its relative contribution is more than twice as large in the
so-called boosted regime, $\pt>150$\,GeV. Clearly, such an effect needs
to be taken into account in experimental analyses of the $\zh$
production process, in particular since it carries such a large
sensitivity to New Physics.

In this paper, we propose a data-driven strategy to extract the
gluon-initiated component (or, more precisely, the non-\dy\ component)
for $\zh$ production. It is based on the comparison of the $\zh$ to the
$\wh$ cross section and the corresponding invariant mass distribution of
the $\vh$ system. The required theory input in the \sm\ is the ratio of
the \dy-like components for $\zh$ and $\wh$ production, which can be
predicted very reliably already now, and is expected to improve even
further in the foreseeable future. We study the impact of various
possible structures in models for New Physics, such as modified Yukawa
couplings, extended Higgs sectors, or vector-like quarks (\vlq{}s).  In
order to estimate the expected experimental uncertainties, we simulate a
recent \abbrev{ATLAS} analysis with Monte-Carlo events, and extrapolate
it to higher luminosities. We find that the estimate of systematic
uncertainties becomes the limiting factor for the measurement,
highlighting the importance of a detailed investigation of systematic
effects, and potentially an optimization of the experimental analysis
towards the extraction of this ratio from data.

%- }}}
%- {{{ ggzh:

\section{Theory prediction for \textit{VH} production}

%- {{{ subsection{Definition and features}

\subsection{Definition and features}\label{sec:definition}

%- {{{ fig:ggzhlo

%
\begin{figure}
  \begin{center}
    \begin{tabular}{cc}
      \includegraphics[height=.15\textheight]{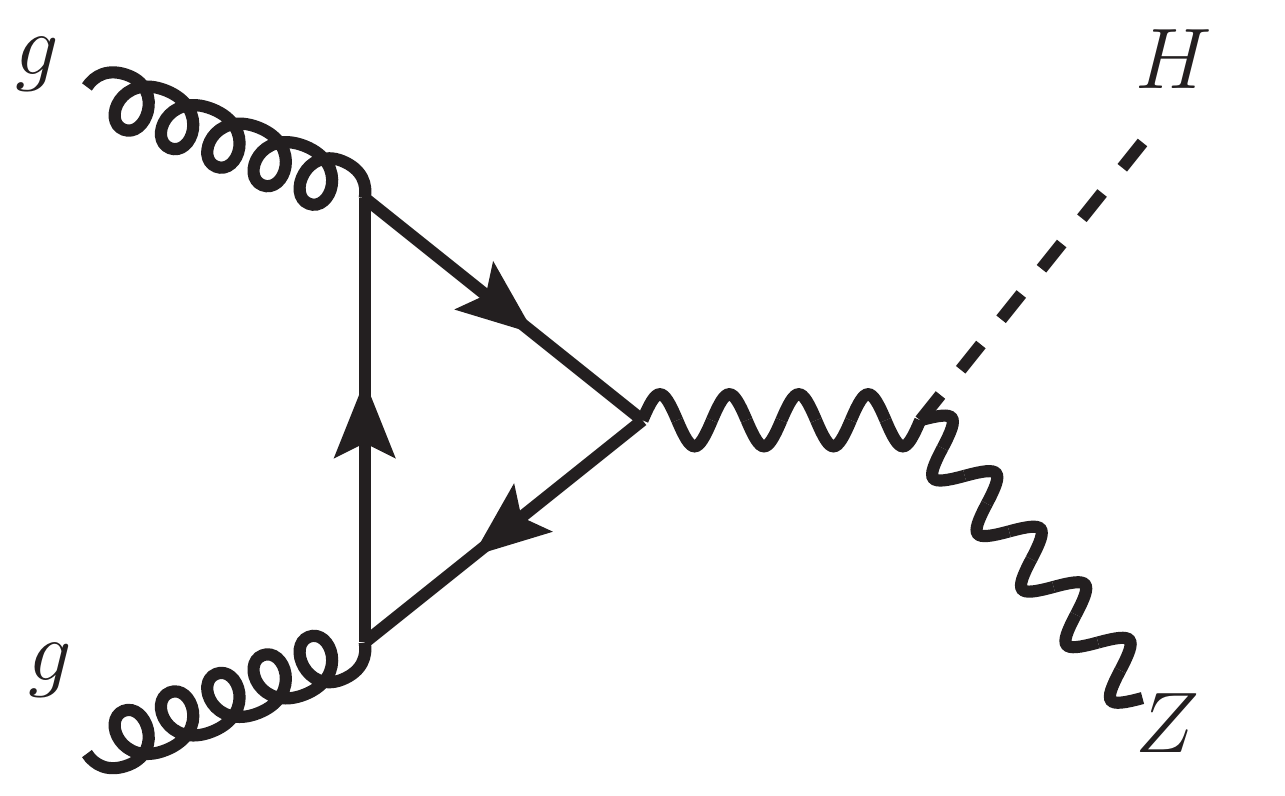}
      &
      \includegraphics[height=.15\textheight]{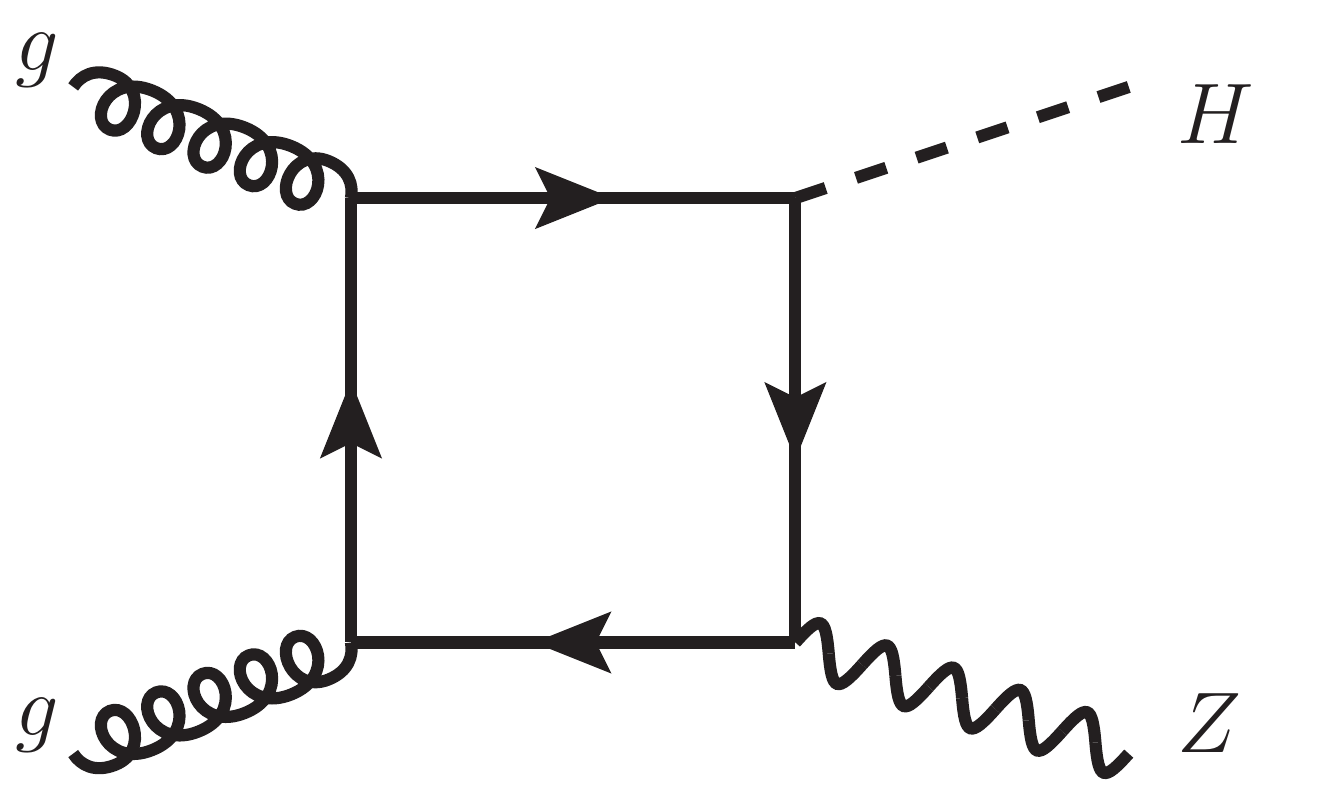}\\ (a)
      & (b)
    \end{tabular}
    \parbox{.9\textwidth}{
      \caption[]{\label{fig:ggzhlo}\sloppy
        Sample Feynman diagrams that contribute to the \ggzh\ process at
        leading order.
        }}
  \end{center}
\end{figure}
%

%- }}}

Let us consider the following theoretical decomposition of the inclusive
$\vh$ production cross section:
\begin{equation}
  \begin{split}
    \sigma^{\vh} = \sigma^{\vh}_{\dy} + \sigma^{\vh}_\text{non-\dy}\,,
    \label{eq:zhdec}
  \end{split}
\end{equation}
where, by definition, the \dy\ component can be written as
\begin{equation}
  \begin{split}
    \sigma^{\vh}_{\dy} =
    \int\dd q^2\,\sigma_{V}(q^2)\,\frac{\dd\Gamma_{V^*\to \vh}}{\dd q^2}
    + \Delta\sigma^{\vh}_\text{EW}\,.
    \label{eq:vhint}
  \end{split}
\end{equation}
In \eqns{eq:zhdec} and \noeqn{eq:vhint}, the electro-weak corrections
$\Delta\sigma^{\vh}_\text{EW}$ are understood to be fully attributed to
$\sigma^{\vh}_{\dy}$, i.e., by definition, $\sigma^{\vh}_\text{non-\dy}$
does not receive any electro-weak corrections. At \lo\ perturbation
theory, the \dy-like terms for $\wh$ are related to those for $\zh$ by
changing external parameters like the gauge boson mass, the gauge
coupling, or the \pdf, all of which can (and are) determined
independently through other processes. The effect of higher orders on
this similarity between the \dy\ components will be studied below.  Note
that any New Physics most likely respects the well-established gauge
symmetry between the $W$ and the $Z$, and thus preserves the strong tie
between the \dy-components for $\wh$ and $\zh$ production. For
example, in a general 2-Higgs-Doublet-Model (\thdm), the ratio of the
\dy\ components for \zh\ and \wh\ production is the same as in the \sm.

Concerning $\sigma^{\vh}_\text{non-\dy}$, the dominant contribution in
the \sm\ for $V=Z$ is due to the gluon-initiated process \ggzh, denoted
by $\sigma_{gg}$. The generic set of diagrams contributing to this
sub-process at \lo\ is shown in \fig{fig:ggzhlo}.  We stress that,
within \qcd{}, $\sigma_{gg}$ is well-defined since it is separately
finite and gauge invariant to all orders of perturbation theory. In \bsm\
theories, also $b\bar b$-initiated contributions may become important in
$\sigma^{\zh}_\text{non-\dy}$. None of these have a correspondence in
$\wh$ production; in fact, in this paper we will assume that only
\zh\ production receives non-\dy\ contributions,
i.e.\ $\sigma^{\wh}_\text{non-\dy}=0$.

The current theoretical precision is quite different for the first two
components on the l.h.s.\ of \eqn{eq:zhdec}. While $\sigma^{\vh}_{\dy}$
is known through
\nnlo\ \qcd\,\cite{Brein:2003wg,Hamberg:1990np,Ferrera:2011bk,
  Ferrera:2014lca,Campbell:2016jau}, i.e.\ $\order{\as^2}$, the current
theory prediction for the total inclusive cross section of $\sigma_{gg}$
is based on the full \lo\ calculation, which is also of order
$\as^2$\,\cite{Kniehl:1990iva,Dicus:1988yh}. At this order,
$\sigma_{gg}$ amounts to about 6\% of the total $\zh$ cross section for
$\mhiggs=125$\,GeV in $pp$ collisions at $13$\,TeV. A full calculation
of the relevant \nlo\ corrections, i.e.\ $\order{\alpha_s^3}$, is not
yet available. However, assuming that it depends only weakly on the
top-quark mass, as it is the case for the gluon-fusion process $gg\to
H$, the \nlo\ correction factor has been found to be of the order of
two, which increases the \ggzh\ contribution to the total cross section
accordingly\,\cite{Altenkamp:2012sx,Harlander:2014wda}. Higher order
terms in $1/\mtop$ were evaluated in \citere{Hasselhuhn:2016rqt}, but
their validity is restricted to an invariant mass $\mzh$ of the
\zh\ system of $\mzh<2\mtop$. Concerning differential distributions, the
amplitudes for 2- and 3-parton final states including the full
quark-mass dependence have been merged in order to obtain a reliable
prediction at large transverse momenta of the Higgs
boson\,\cite{Hespel:2015zea,Goncalves:2015mfa}. For
$\sigma_{\dy}^{\vh}$, also electro-weak corrections are
known\,\cite{Ciccolini:2003jy,Denner:2011id,Granata:2017iod}, while
  they are unavailable for $\sigma_{gg}$ at the time of this writing.
  As a consequence, the estimated theoretical accuracy due to scale
  variation for the \dy-like component is at the sub-percent level,
  while it reaches up to about 25\% for $\sigma_{gg}$ at \nlo. Including
  \nll\ resummation, this reduces to about
  7\%\,\cite{Harlander:2014wda}. The \pdf{} uncertainties\footnote{Using
    \texttt{PDF4LHC15\_nnlo\_100}\,\cite{Butterworth:2015oua}.} are at
  the 2\% and 4\% level for the \dy\ and the $gg$ component,
  respectively. \nnlo+\abbrev{PS} implementations of the \wh\ and the
  \zh\ process have been presented in
  \citeres{Astill:2016hpa,Astill:2018ivh}.

%- }}}
%- {{{ subsection{New physics effects}

\subsection{New-Physics effects}

The gluon-initiated component reveals some interesting features which
predestines it as a probe for New Physics. First of all, it is
loop-induced, which means that it is particularly sensitive to as-of-yet
unknown particles which might couple the initial-state gluons to the
$\zh$ final state. Second, the dominant contribution in the \sm{} is due
to top-quark loops, which lead to a characteristic threshold-structure
in various kinematical distributions of the cross section. The
application of appropriate cuts thus allows for enriching the $\zh$
sample with gluon-initiated events, as pointed out in
\citere{Englert:2013vua}. Through the box diagrams,
\fig{fig:ggzhlo}\,(b), the cross section also receives a dependence on
the top-quark Yukawa coupling, which is amplified by the fact that the
box diagrams interfere destructively with the triangle diagrams,
\fig{fig:ggzhlo}\,(b). Another interesting feature which appears in many
\bsm\ models are $s$-channel contributions due to additional
Higgs bosons\,\cite{Harlander:2013mla}. They either add to the
triangle-component of $\sigma_{gg}$, or they occur in the process $b\bar
b\to \zh$. For future reference, we refer to the latter contribution as
$\sigma_{b\bar b}$, distinguishing it from the $b\bar b$-contribution to
$\sigma_{\dy}^{\vh}$ by requiring that $\sigma_{b\bar b}= 0$ in the
limit of a vanishing bottom-quark Yukawa coupling. In the \sm, this
contribution amounts to less than 0.1\% of the \dy\ term.

Many of such New-Physics effects on $\sigma_{gg}$ as well as
$\sigma_{b\bar b}$ can be investigated with the help of the program
\vhnnlo\,\cite{Brein:2012ne,Harlander:2018yio}.

%- {{{ fig:ptbsm:

%
\begin{figure}
  \begin{center}
    \begin{tabular}{cc}
      \includegraphics[height=.33\textheight]{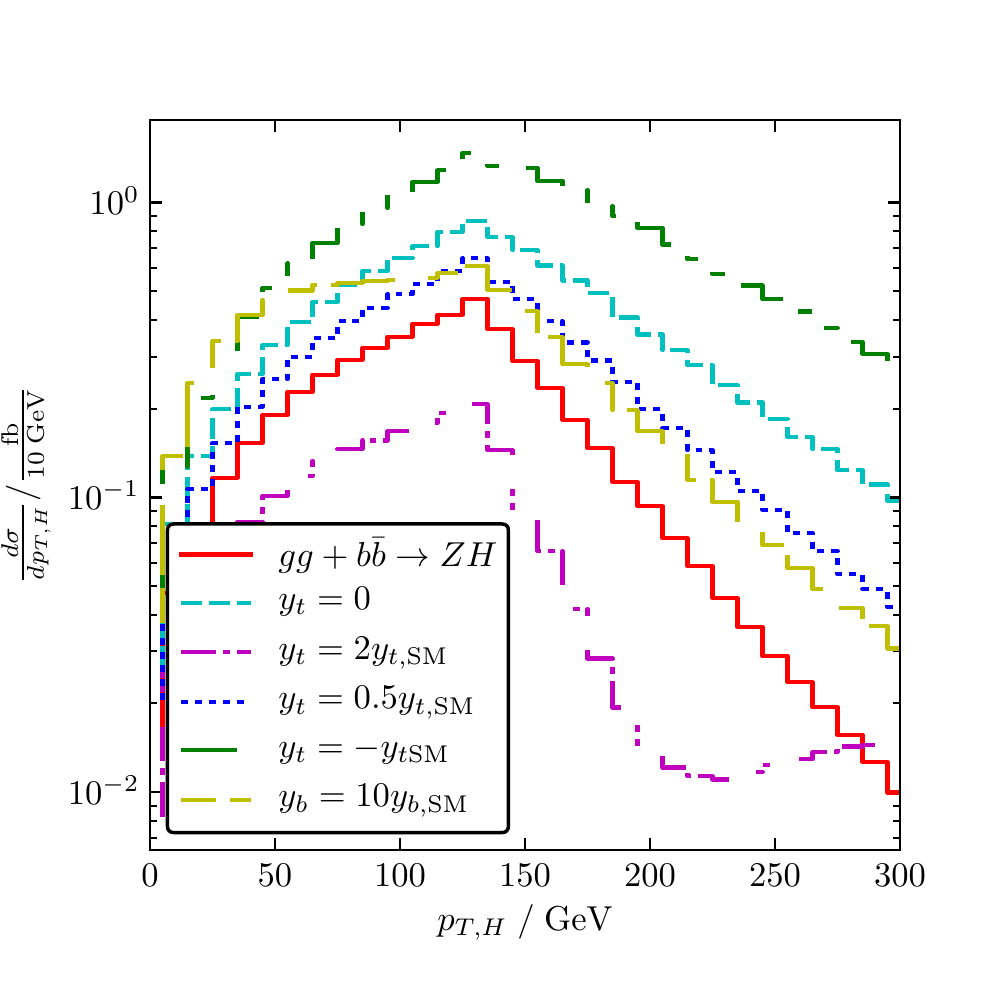} &
      \includegraphics[height=.33\textheight]{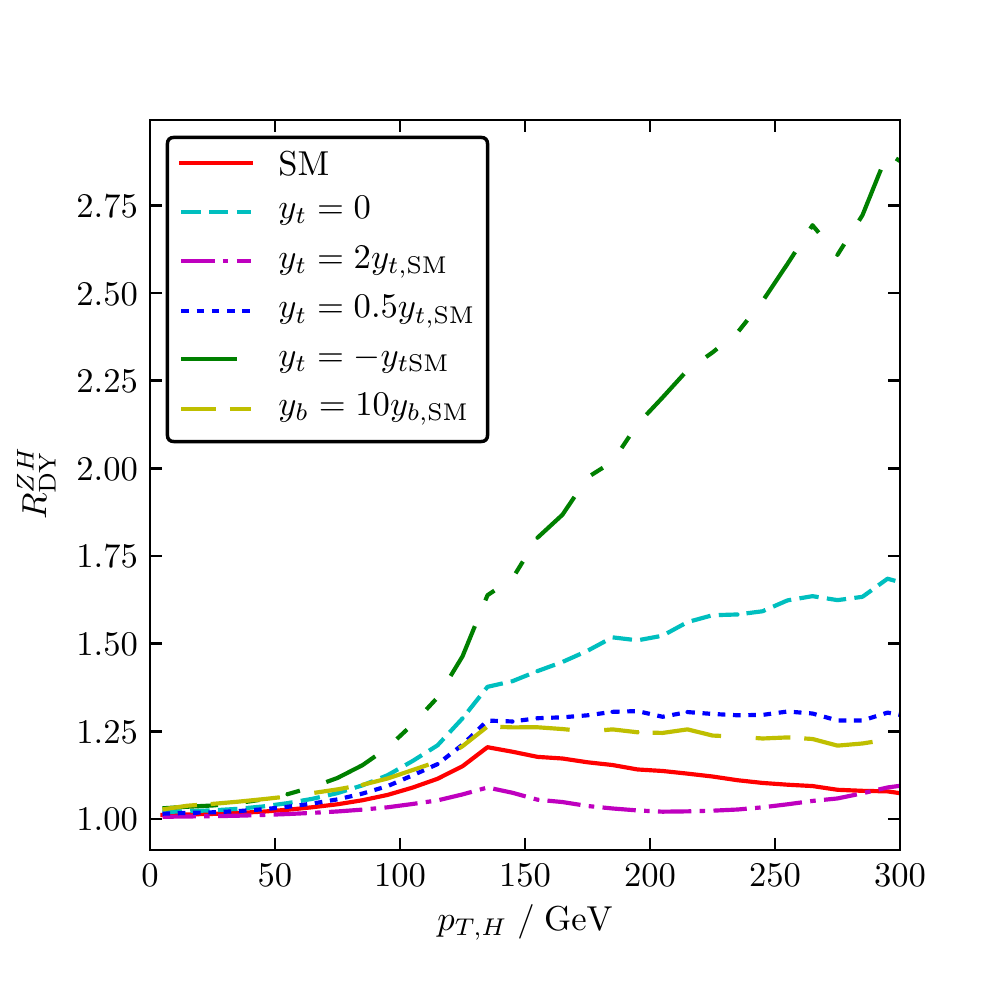}\\
      (a) & (b)
    \end{tabular}
    \parbox{.9\textwidth}{
      \caption[]{\label{fig:ptbsm}\sloppy (a) The $\pt$ spectrum of the
        Higgs boson produced through the $gg$- and $b\bar b$-processes
        for different values of the top- and bottom-quark Yukawa
        couplings. (b) The ratio of the full \zh\ cross section to the
        \dy\ component. The latter was obtained at \nnlo\ with the help
        of \mcfm\,\cite{Campbell:2016jau}, the New-Physics effects in
        \ggzh\ were calculated at \lo\ (i.e.\ $\order{\alpha_s^2}$) with
        \vhnnlo\,\cite{Brein:2012ne,Harlander:2018yio}, using
        \texttt{PDF4LHC15\_nnlo} \pdf{}s with $\alpha_s(M_Z)=0.118$ in
        both cases~\cite{Butterworth:2015oua}. The (local) minimum at
        $\pt\sim 230$\,GeV for $y_t=2y_{t,\text{\sm}}$ is an effect from
        the box-triangle interference. }}
  \end{center}
\end{figure}
%

%- }}}

Deviations from the \sm, like modified Yukawa couplings, new colored
particles, or an extended Higgs sector, are thus likely to manifest
themselves in the $\zh$ final state through the gluon- or $b\bar
b$-initiated component of the cross section, given that one considers a
suitable observable.  In \sct{sec:extract} we will argue that the
\dy\ and the non-\dy\ $\zh$ contribution can be isolated to a high
degree by considering the ratio\footnote{ Here and in what follows, we
  use the notation in \eqn{eq:RDY} for a generic distribution. More
  explicitly, we may write
\begin{equation}
  \begin{split}
    R^{\zh}_{\dy}(x) = \frac{\dd\sigma^{\zh}/\dd
      x}{\dd\sigma^{\zh}_{\dy}/\dd x}\,.
    \label{eq:Rdyx}
  \end{split}
\end{equation}
for a distribution in a specific kinematic variable $x\in\{\pth,
\mvh,\ldots \}$.}
\begin{equation}
  \begin{split}
    R^{\zh}_{\dy} \equiv \frac{\sigma^{\zh}}{\sigma^{\zh}_{\dy}}\,.
    \label{eq:RDY}
  \end{split}
\end{equation}
An obvious kinematical parameter to consider would be the transverse
momentum of the Higgs boson. Indeed, as shown in \fig{fig:ptbsm}, the
$p_T$ distribution of the Higgs boson produced in non-\dy\ processes
exhibits a significant dependence on New Physics (a non-\sm\ Yukawa
coupling in this case).

%- {{{ fig:mzhbsm:

%
\begin{figure}
  \begin{center}
    \begin{tabular}{cc}
      \includegraphics[height=0.35\textheight]{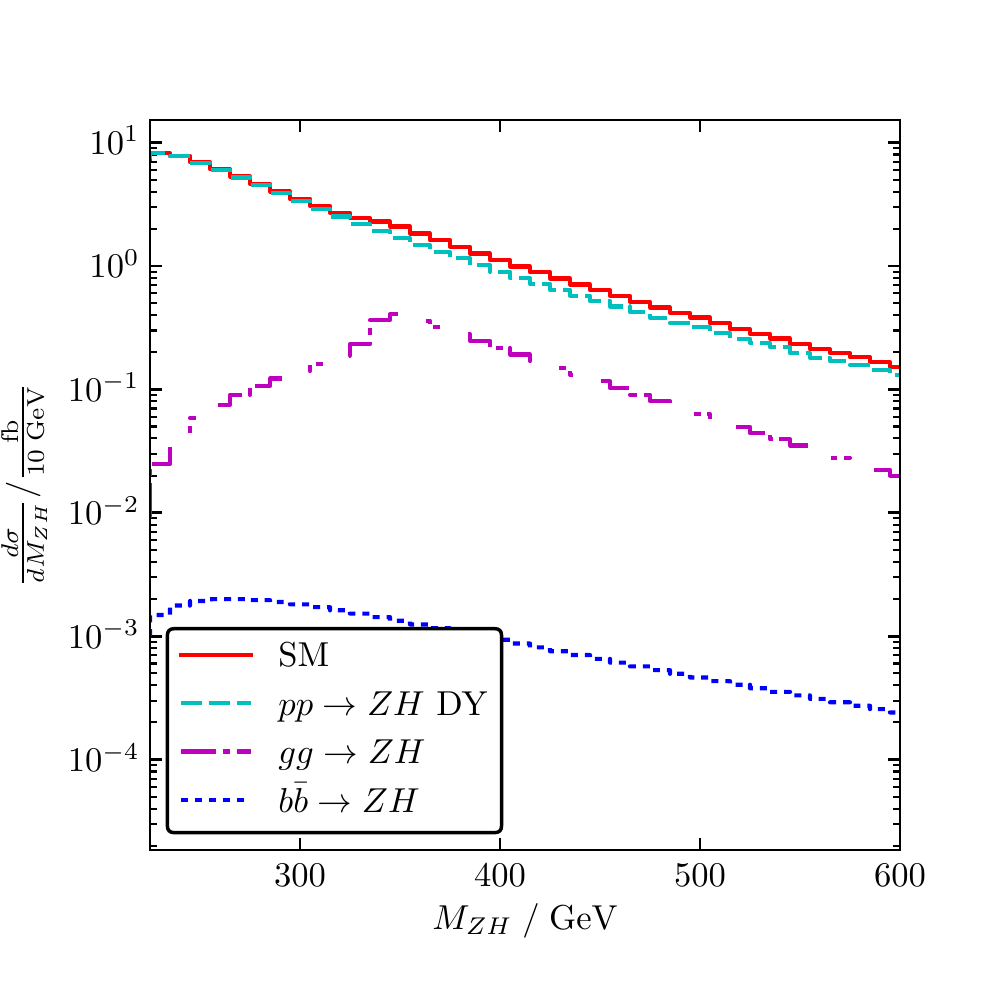} &
      \includegraphics[height=.35\textheight]{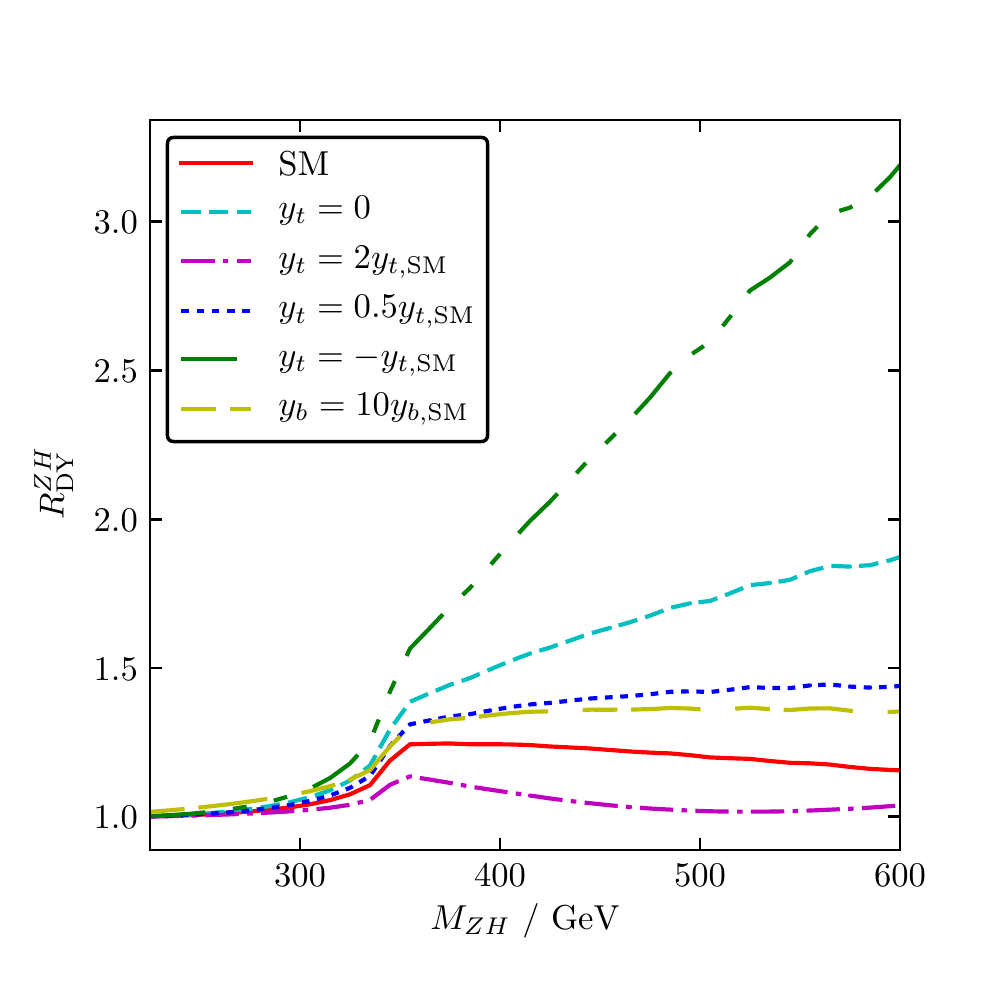}
      \\ (a) & (b)\\ \includegraphics[height=.35\textheight]{%
        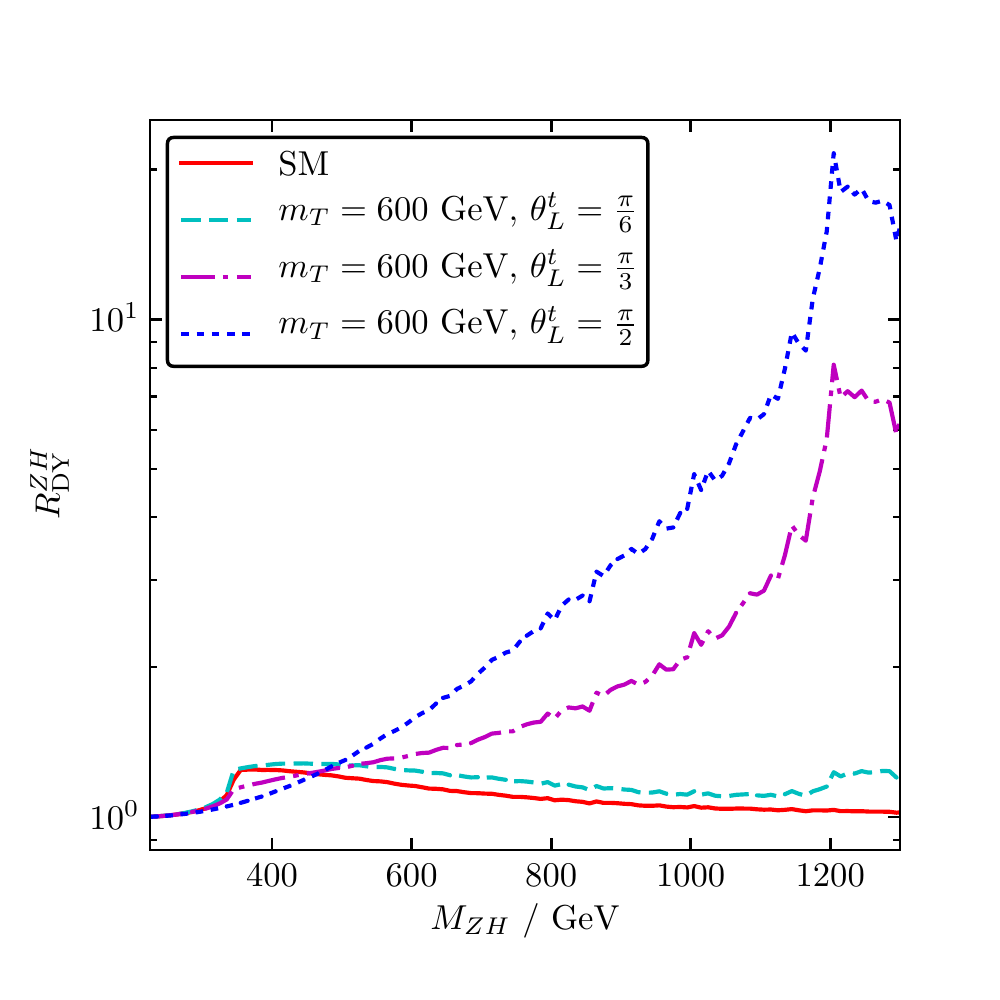} &
      \includegraphics[height=.35\textheight]{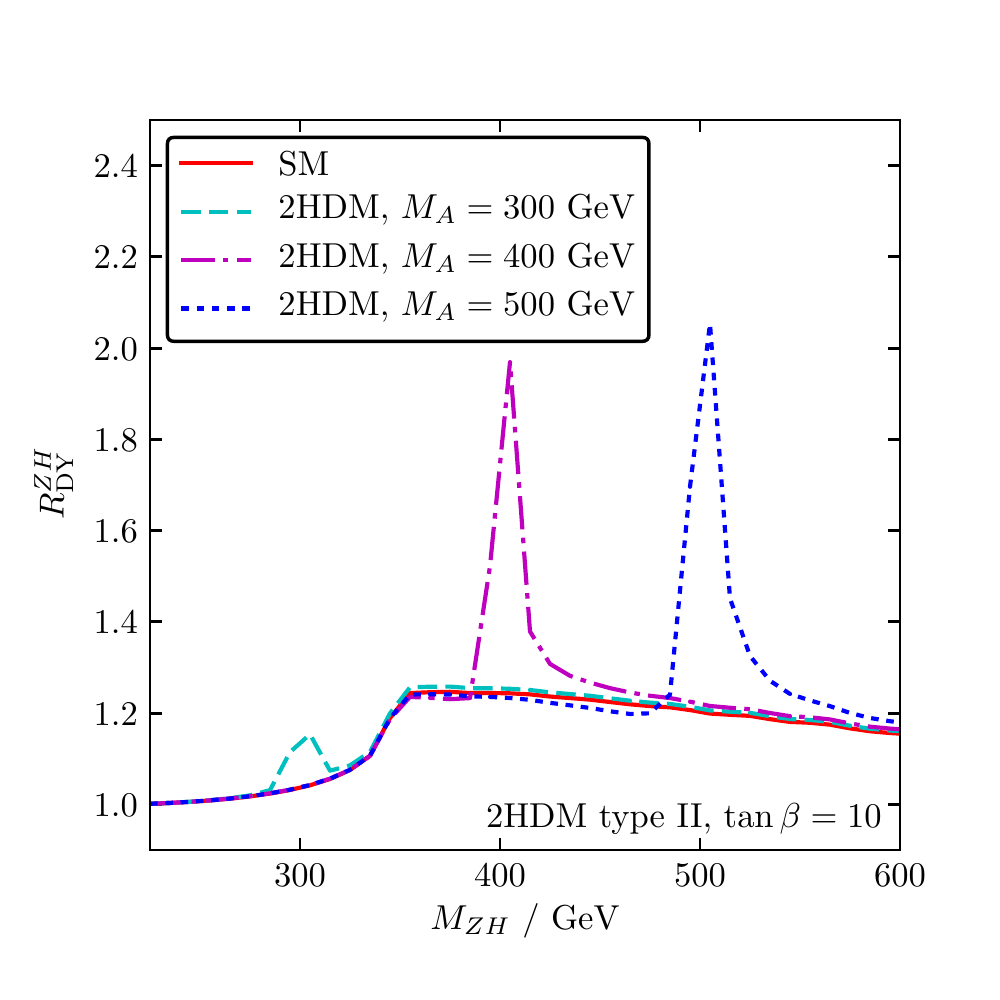}
      \\ (c) & (d)
    \end{tabular}
    \parbox{.9\textwidth}{
      \caption[]{\label{fig:mzhbsm} \sloppy (a) The $\zh$ invariant mass
        spectrum in the \sm: total cross section (solid red),
        \dy\ component (dashed cyan), \ggzh\ component (dash-dotted
        purple), \bbzh\ component (dotted blue). (b)~The ratio
        $R_{\dy}^{\zh}$ of the full $\zh$ cross section to the
        \dy\ component in the \sm\ (solid red) and for modified values
        of the top- and bottom-quark Yukawa coupling, including various
        New-Physics effects. (c)~$R_{\dy}^{\zh}$ for the \sm+\vlq{}s of
        mass 600\,GeV, and various values of the \vlq\ mixing
        angle. (d)~$R_{\dy}^{\zh}$ for the production of a light
        \sm-like Higgs in the \thdm\ with a pseudo-scalar of different
        masses including $\bbzh$. The \dy-like component was obtained at
        \nnlo\ with the help of \mcfm\,\cite{Campbell:2016jau}, the
        New-Physics effects in \ggzh\ were calculated at
        \lo\ (i.e.\ $\order{\alpha_s^2}$) with
        \vhnnlo\,\cite{Brein:2012ne,Harlander:2018yio}, using
        \texttt{PDF4LHC15\_nnlo} \pdf{}s with $\alpha_s(M_Z)=0.118$ in
        both cases~\cite{Butterworth:2015oua}.}}
  \end{center}
\end{figure}
%

%- }}}

In this paper, however, we want to focus on the invariant mass
distribution $\mzh$ of the $\zh$ system, since we find that it reveals
particularly distinct features that allow to identify various
New-Physics models, especially when normalized to the \dy-like $\zh$
contribution. Examples are shown in \fig{fig:mzhbsm}, which include the
effect of both $\ggzh$ and $\bbzh$, the latter of which becomes relevant
in scenarios with enhanced bottom-quark Yukawa coupling. Experimentally,
the invariant mass for the $\mvh$ system may be difficult to access, and
other observables such as the $\pt$ spectrum may be more
advantageous. The optimal observable is best determined within an
experimental analysis where all the systematic uncertainties are
available. The reconstruction of the \wh\ invariant mass for our
numerical simulation is outlined in \sct{sec:numerics}. The general
idea of the current paper is independent of the observable under
consideration.

Throughout the paper, we set
\begin{equation}
  \begin{split}
    \sqrt{s}=13\,\text{TeV}\,,\qquad
    \mtop=173\,\text{GeV}\,,\qquad \mhiggs=125\,\text{GeV}\,,
%    \label{eq:}
  \end{split}
\end{equation}
unless indicated otherwise. As already pointed out in
\citere{Englert:2013vua}, the contribution of \ggzh\ to the total cross
section is typically rather small in the kinematical region below the
top-quark threshold. The distribution above $2\mtop$, on the other hand,
distinctly reflects the impact of New Physics. Specifically, this region
crucially depends on the top-quark Yukawa coupling, both in magnitude
and sign, as shown in \fig{fig:mzhbsm}\,(a) and (b). In addition, new
heavy particles which contribute to the effective $ggZH$ coupling might
also reveal extra threshold structures in the invariant mass spectrum,
as shown using the example of vector-like quarks in
\fig{fig:mzhbsm}\,(c). Non-minimal Higgs bosons which contribute through
$s$-channel exchange lead to yet other features in this spectrum, see
\fig{fig:mzhbsm}\,(d), which shows $R_\mathrm{DY}^{\zh}$ for a \thdm.
The peak structure is dominated by the $\bbzh$ process in this case (see
also \citere{Harlander:2018yio}).

%- }}}

%- }}}
%- {{{ section{Extracting \ggzh from data}

\section{Extracting the non-\dy\ component from data}\label{sec:extract}

%- {{{ The double ratio:

\subsection{The double ratio}\label{sec:double_ratio}

The high accuracy to which the \dy\ component is known theoretically
suggests a simple comparison of the experimentally determined $\vh$ rate
to the theoretical prediction of its \dy\ component, and thus the
extraction of the non-\dy\ to the \dy\ ratio directly from
$R_{\dy}^{\zh}=\sigma^{\zh}/\sigma^{\zh}_{\dy}$ of \eqn{eq:RDY}:
\begin{equation}
  \begin{split}
    \frac{\sigma^{\zh}_\text{non-\dy}}{\sigma_\dy^{\zh}} =
    R_{\dy}^{\zh}-1 = 
    \frac{\sigma^{\zh}}{\sigma_{\dy}^{\zh}} - 1\,,
    \label{eq:Rdirect}
  \end{split}
\end{equation}
with the \dy-like cross section, $\sigma_\dy^{\zh}$, taken from theory, and
the full $\zh$ cross section $\sigma^{\zh}$ from experiment. Such an
experiment/theory comparison suffers from potential systematic
uncertainties though, due to detector simulation, unfolding, and the
like.

In this paper, we propose to analyze the data from Higgs-Strahlung by
making use of a very specific feature for this process which has been
alluded to in \sct{sec:definition}, namely the similarity between the
$\zh$ and the $\wh$ process.\footnote{We note that, at the level of the
  actual Drell-Yan process of virtual $V$ production, $pp\to V^\ast$,
  the symmetry between $V=W$ and $V=Z$ has been used before as an
  alternative way to measure the $W$ boson mass at hadron
  colliders\,\cite{Giele:1998uh}.}  For this purpose, let us define the
double ratio
\begin{equation}
  \begin{split}
    R_R^{\zw} =
    \frac{\sigma^{\zh}/\sigma^{\wh}}{\sigma^{\zh}_{\dy}/\sigma^{\wh}}
    \equiv \frac{R^{\zw}}{R_{\dy}^{\zw}}\,.
    \label{eq:RR}
  \end{split}
\end{equation}
Obviously, if all quantities are evaluated theoretically, it is
$R_R^{\zw} = R^{\zh}_{\dy}$, cf.\ \eqn{eq:RDY}. Here, however, we
suggest to measure the numerator $R^{\zw}=\sigma^{\zh}/\sigma^{\wh}$ of
the double ratio in \eqn{eq:RR} from experimental data. Despite the different
final states for $\zh$ and $\wh$ production, we expect that a number of
systematic experimental uncertainties cancel, in particular if the
parameters of the analyses for $\zh$ and $\wh$ are aligned as much as
possible. A rough estimate of the experimental uncertainty will be described below.

The denominator of \eqn{eq:RR}, on the other hand, referred to as the
\dy\ ratio in what follows, can be calculated within the \sm\ with
rather high precision, as will be discussed below. In addition, it can
hardly be affected by any New-Physics effects, because of the strong
theoretical and experimental constraints on the electro-weak gauge
couplings, as already discussed in \sct{sec:definition}.

We note that the comparison of $\wh$ to $\zh$ as a probe for New Physics
has been first suggested in \citere{Harlander:2013mla}, where the
\thdm\ was considered as an example at the level of total cross
sections, partly with boosted topology. In this paper we provide a much
more elaborate investigation of that proposal, on the basis of
differential quantities and including an estimate of the expected
experimental uncertainty through the analysis of a simulated event sample.

%- }}}
%- {{{ subsection{Theory prediction for the \dy\ ratio}

\subsection{Theory prediction for the \dy\ ratio}\label{sec:theorydy}

At the level of the total cross section, $\RZWDY{}{}$ receives corrections of only 0.2\% at \nlo, while the \nnlo\ corrections on
top of that are at the per-mill level. This is quite remarkable as the
\nlo\ corrections on the numerator and denominator in that ratio amount
to 16\%; the \nnlo\ corrections on the other hand, are less than 1\% on
top of that.

\begin{figure}
  \begin{center}
    \begin{tabular}{cc}
      \parbox{.45\textwidth}{%
        \begin{tabular}{c}
          \includegraphics[viewport=30 30 300
            210,width=.45\textwidth]{%
            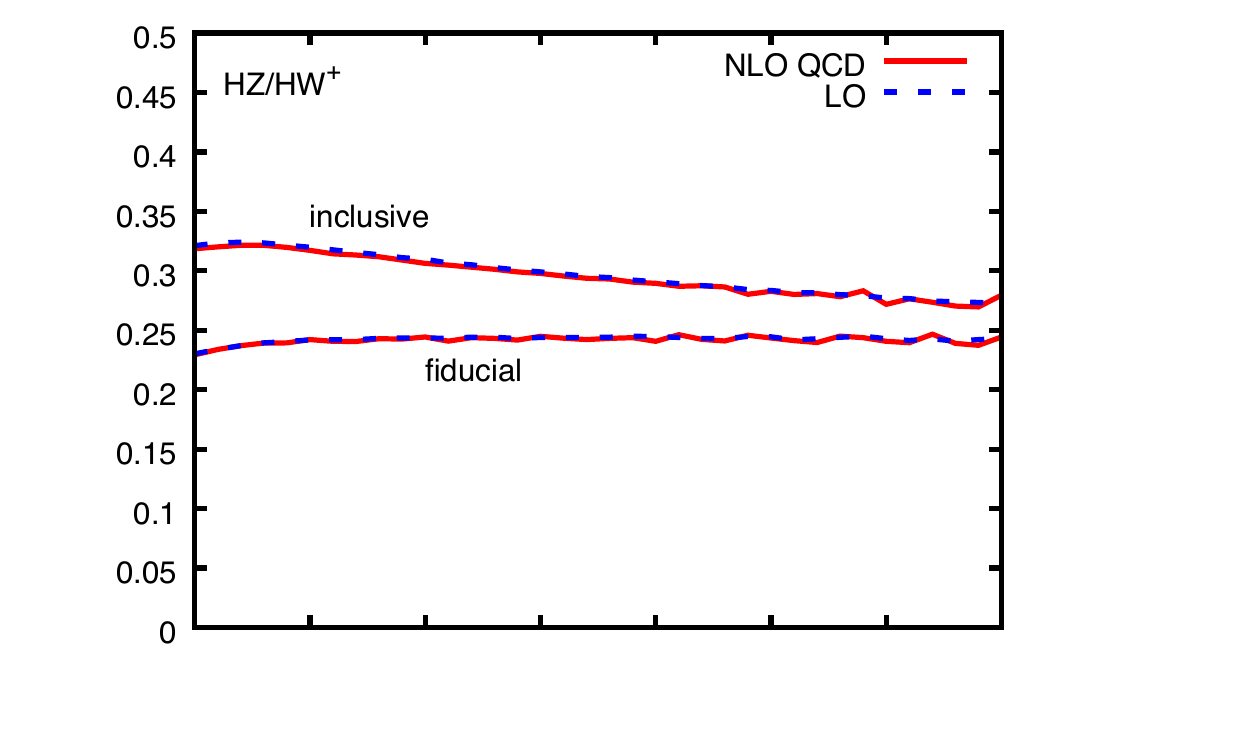}\\[0em]
          \includegraphics[viewport=30 30 300 100,width=.45\textwidth]{%
            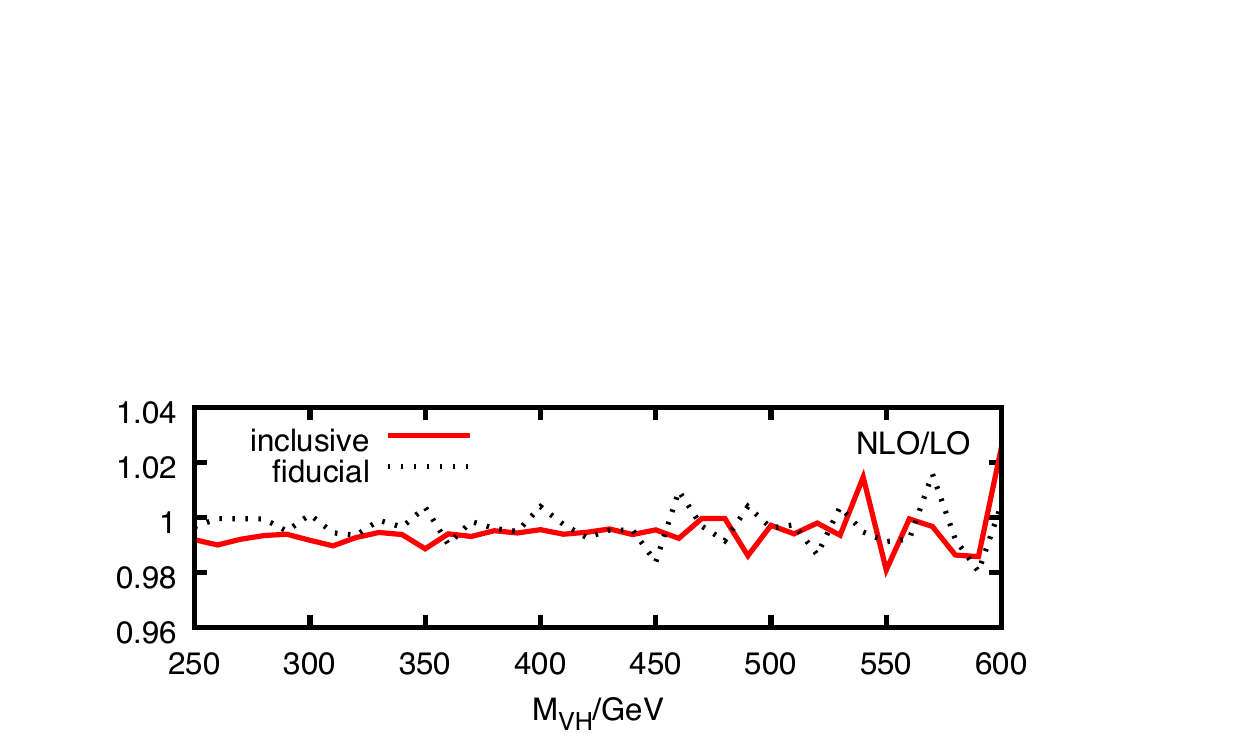}
        \end{tabular}}
      &
      \parbox{.45\textwidth}{%
        \begin{tabular}{c}
          \includegraphics[viewport=30 30 300
            210,width=.45\textwidth]{%
            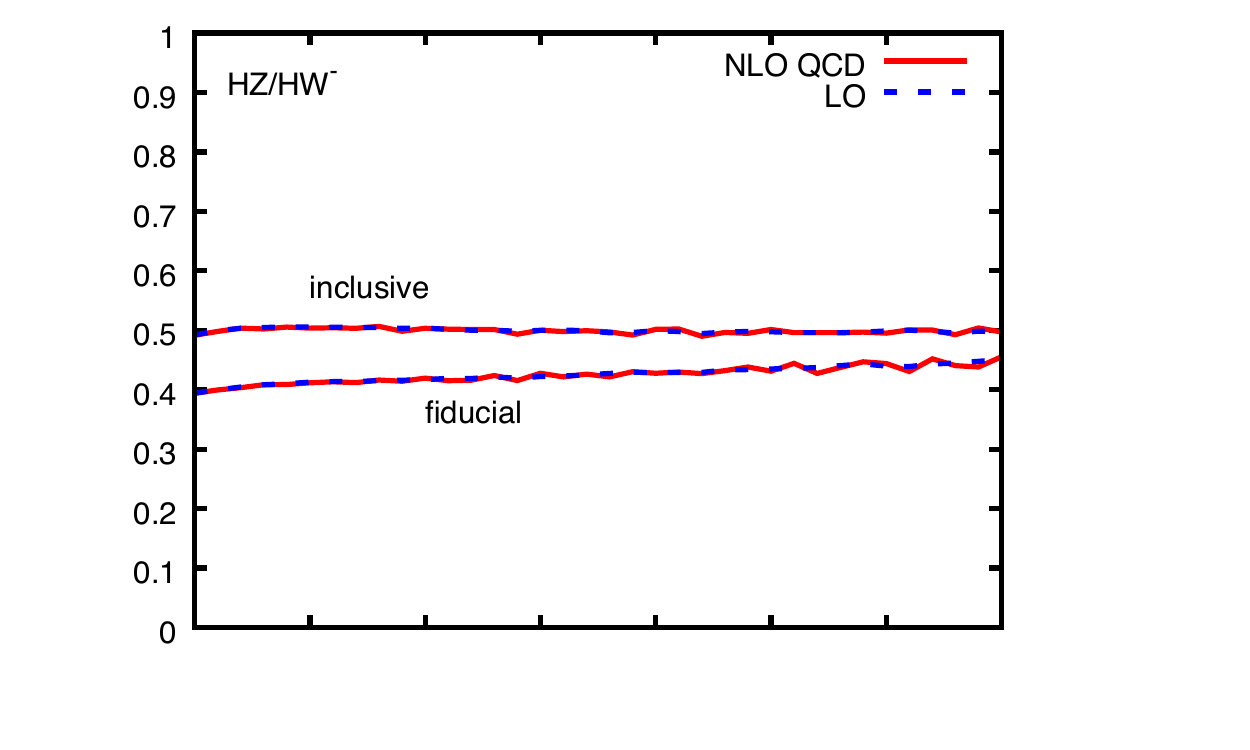}\\[0em]
          \includegraphics[viewport=30 30 300 100,width=.45\textwidth]{%
            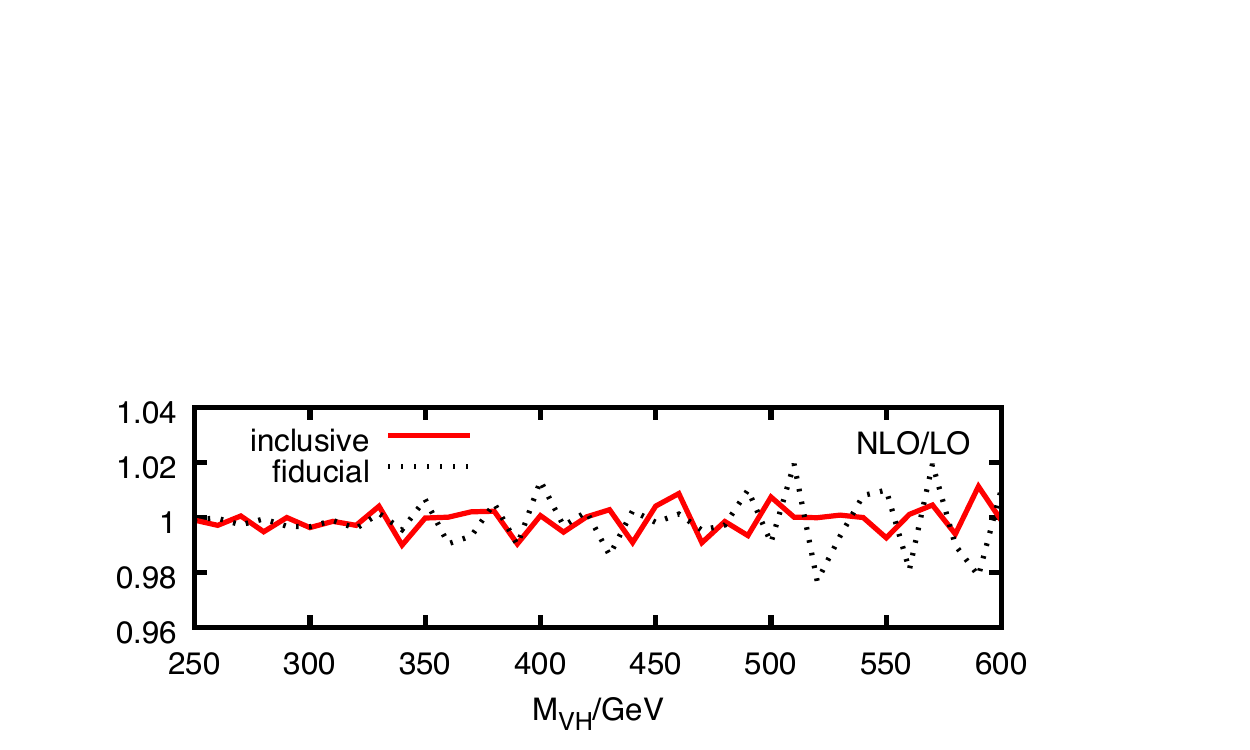}
        \end{tabular}}\\[9em]
      (a) & (b)
    \end{tabular}
    \parbox{.9\textwidth}{
      \caption[]{\label{fig:zw_nloqcd}\sloppy \qcd\ corrections
        (including $qg\to q\vh$) to the ratio $R^{\zw}_{\dy}$ for (a)
        $W=W^+$ and (b) $W=W^-$ as a function of the $\vh$ invariant
        mass, $V\in \{Z,W\}$. The dashed/solid line in the upper parts
        of the plots show the \lo/\nlo\ \qcd\ result, the lower parts
        show the ratio of the two. Obtained with
        \hawk\,\cite{Denner:2011id,Denner:2014cla} (only the decays
        $Z\to l^+l^-$ and $W\to l\nu$ are included) using
        \texttt{NNPDF23\_qed\_nlo} \pdf{}s with
        $\alpha_s(M_Z)=0.118$\,\cite{Ball:2012cx}.}}
  \end{center}
\end{figure}

As a function of \mvh, the \nlo\ corrections on the \dy-ratio are at or
below the 1\% level, as shown in \fig{fig:zw_nloqcd}.  This holds for
both the fully inclusive as well as the ``fiducial'' cross section,
where the latter is evaluated according to \citere{deFlorian:2016spz} by
applying the following cuts:
\begin{equation}
  \begin{split}
    p_T^l> 15\,\text{GeV}\,,\qquad
    y_l<2.5\,,\qquad
    75\,\text{GeV} < m_{ll} <105\,\text{GeV}\,,
%    \label{eq:}
  \end{split}
\end{equation}
where $p_T^l$ and $y_l$ is the transverse momentum and rapidity of a
charged lepton, respectively, and $m_{ll}$ is the invariant mass of a
charged lepton pair (the latter cut only applies to $\zh$ production, of
course).  Using \mcfm, we have also checked that the \nnlo\ corrections on
the \dy-ratio are negligible for all relevant values of $\mvh$. For the
\nlo\ prediction, we thus estimate the uncertainty due to uncalculated
\qcd\ corrections to be less than 1\%.

\begin{figure}
  \begin{center}
    \begin{tabular}{cc}
      \parbox{.45\textwidth}{%
        \begin{tabular}{c}
          \includegraphics[viewport=30 30 300
            210,width=.45\textwidth]{%
            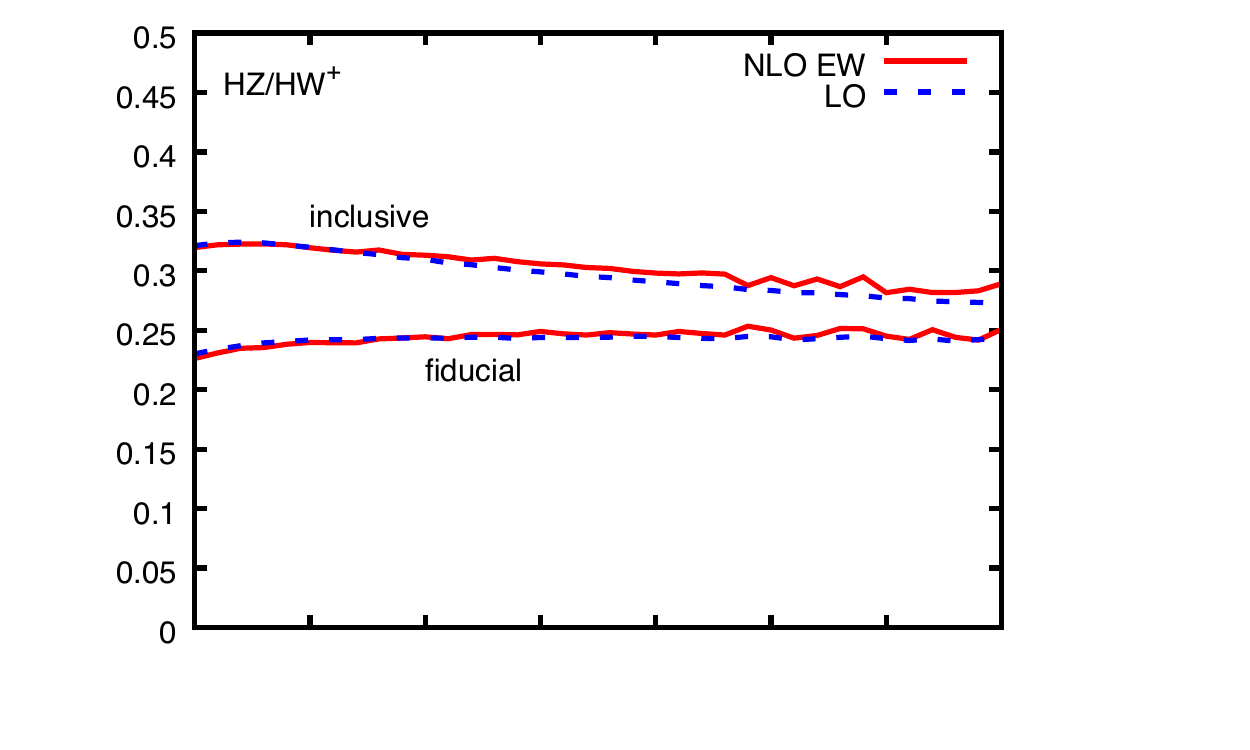}\\[0em]
          \includegraphics[viewport=30 30 300 100,width=.45\textwidth]{%
            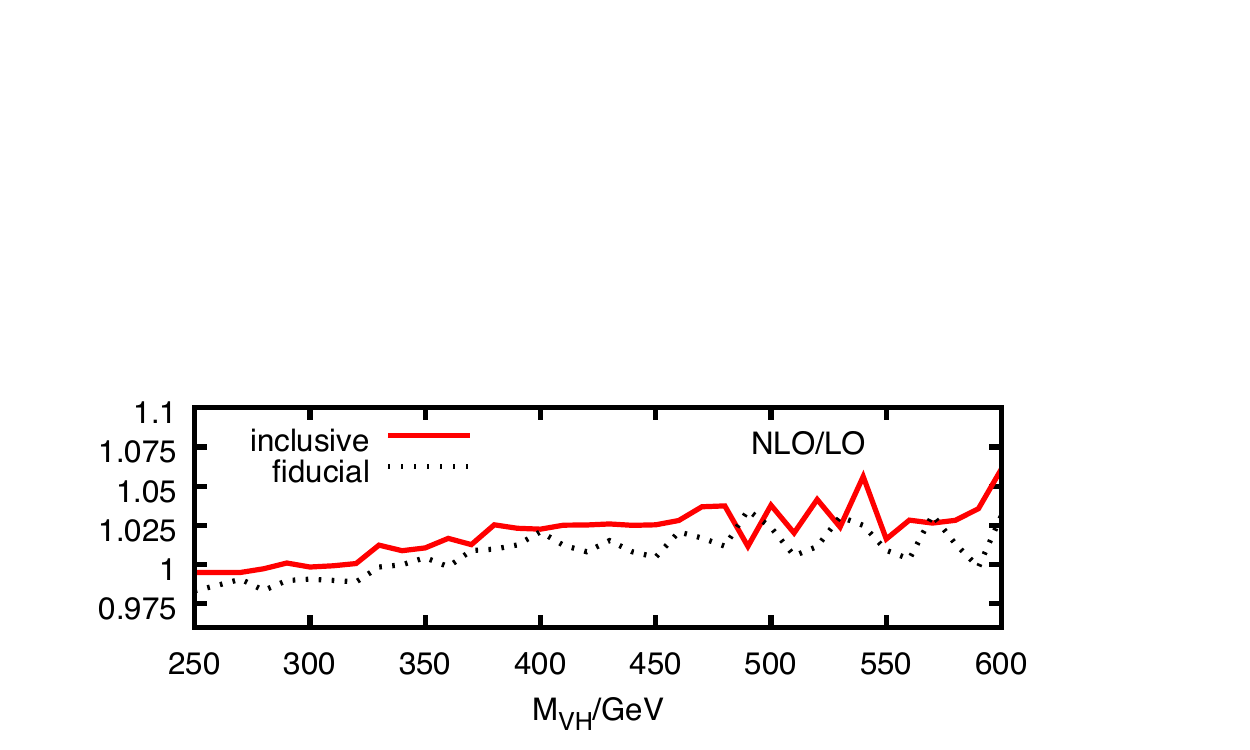}
        \end{tabular}}
      &
      \parbox{.45\textwidth}{%
        \begin{tabular}{c}
          \includegraphics[viewport=30 30 300
            210,width=.45\textwidth]{%
            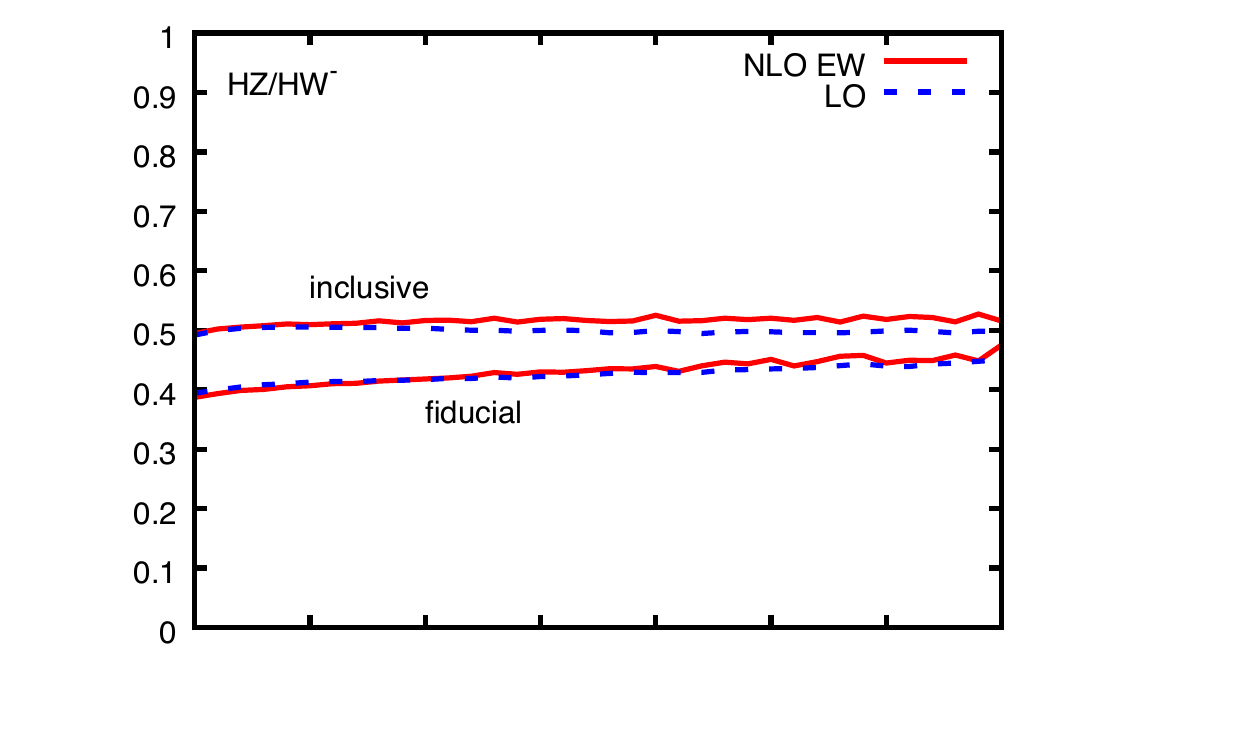}\\[0em]
          \includegraphics[viewport=30 30 300 100,width=.45\textwidth]{%
            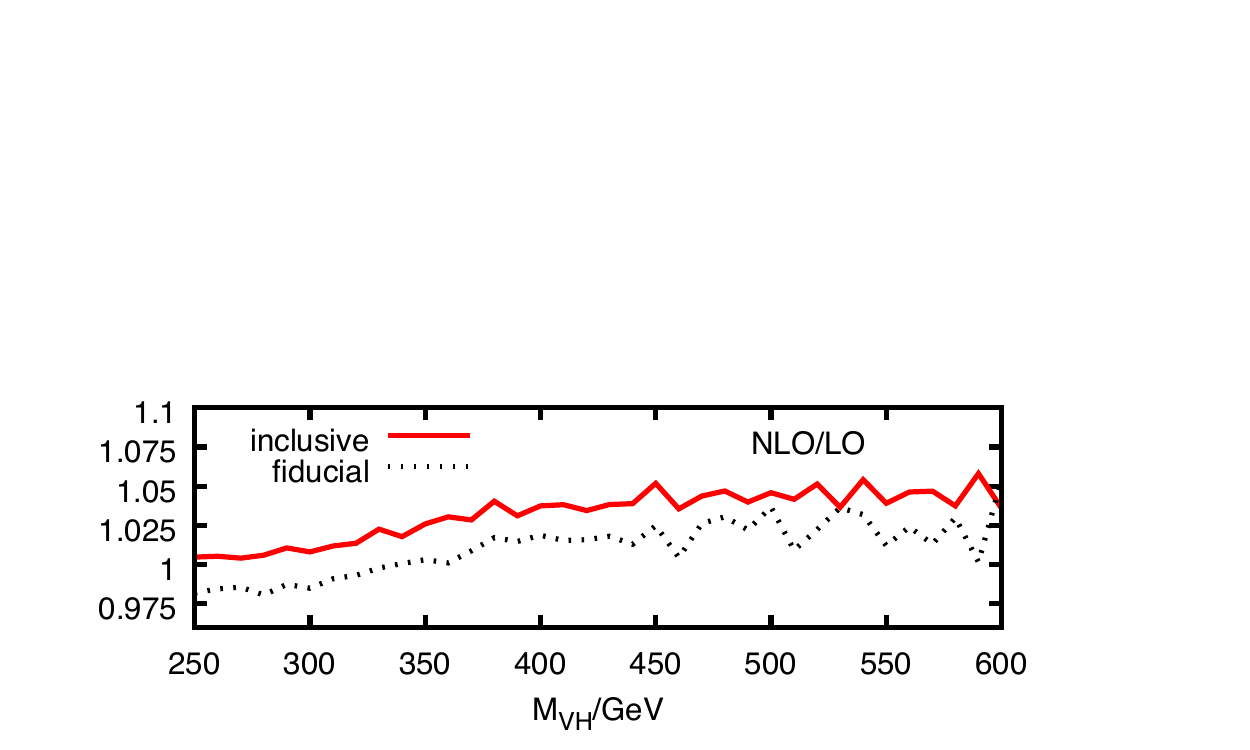}
        \end{tabular}}\\[9em]
      (a) & (b)
    \end{tabular}
    \parbox{.9\textwidth}{
      \caption[]{\label{fig:zw_nloew}\sloppy (a),(b): Same as
        \fig{fig:zw_nloqcd}, but for electro-weak corrections
        (\textit{ex}cluding $\gamma q\to q\vh$).}}
  \end{center}
\end{figure}

Due to the different electric charge of $W$ and $Z$ and their different
decay patterns, one may expect a larger sensitivity of the ratio
$\RZWDY{}{}$ to electro-weak corrections in comparison to the
\qcd\ effects. Indeed, employing
\hawk\,\cite{Denner:2011id,Denner:2014cla} to study these effects, we
find that they amount up to about 5\% on $\RZWDY{}{}$ when the $Z$ decay
into charged leptons is considered,\footnote{We use the default setting
  \texttt{sbarelep=1} of \hawk\ for the final-state leptons, meaning
  that they are not re-combined with photons.  For the $Z$ decay into
  neutrinos, which is not considered in our analysis, the electro-weak
  corrections amount to about 10\%.} see \fig{fig:zw_nloew}. Compared to
the \qcd\ corrections, the electro-weak effects on $R_{\dy}^{\zw}$ show
a stronger dependence on $\mvh$, albeit in a very continuous and
monotonous way.

\begin{figure}
  \begin{center}
    \begin{tabular}{cc}
      \parbox{.45\textwidth}{%
        \begin{tabular}{c}
          \includegraphics[viewport=30 30 300
            210,width=.45\textwidth]{%
            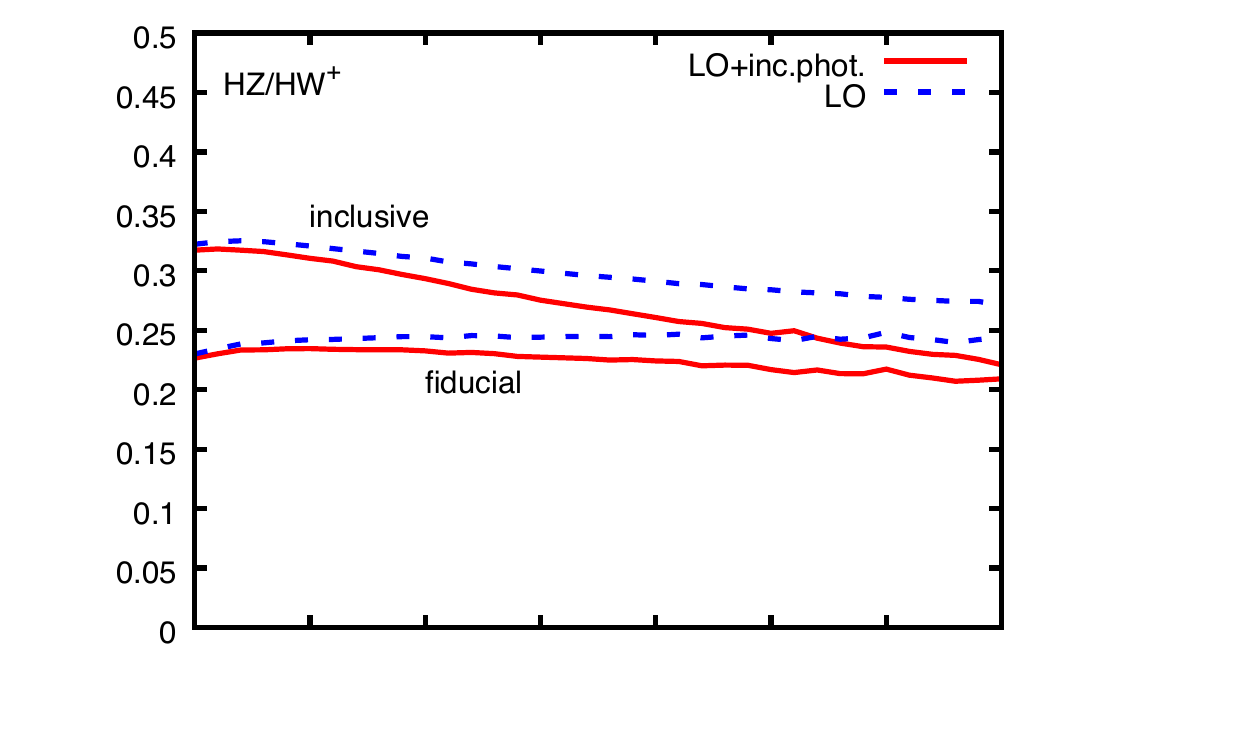}\\[0em]
          \includegraphics[viewport=30 30 300 100,width=.45\textwidth]{%
            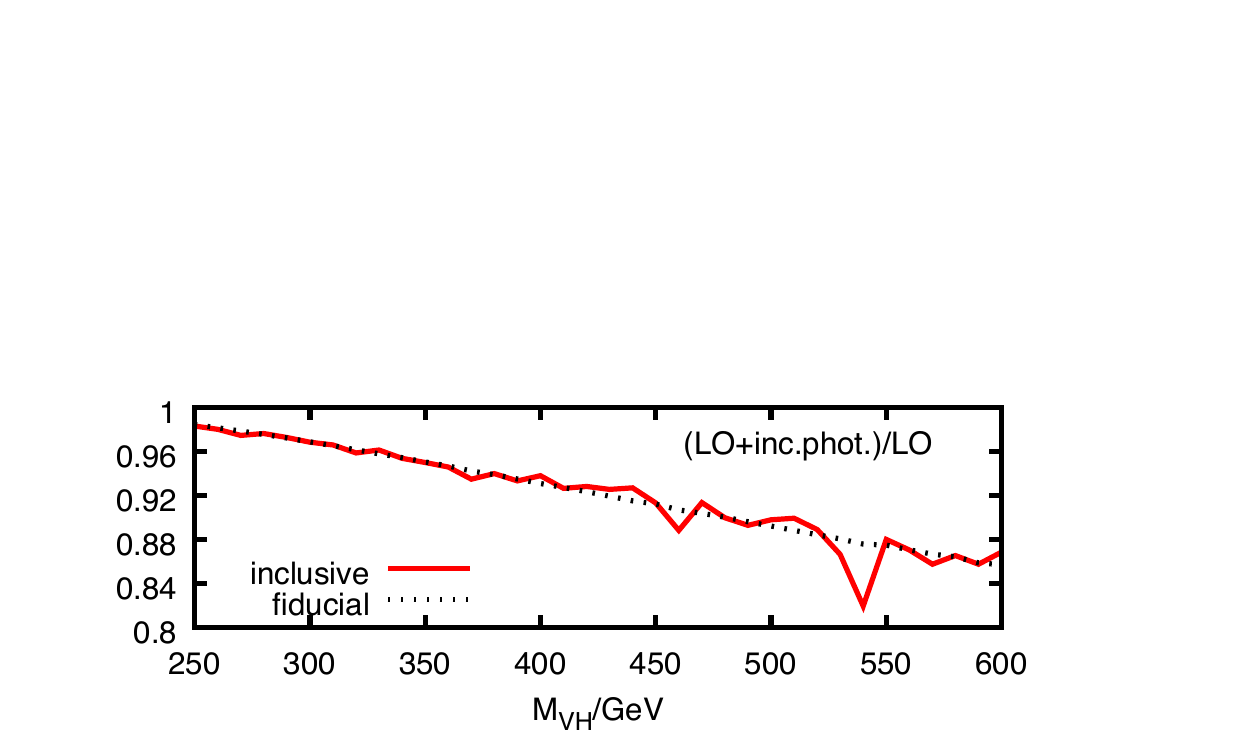}
        \end{tabular}}
      &
      \parbox{.45\textwidth}{%
        \begin{tabular}{c}
          \includegraphics[viewport=30 30 300
            210,width=.45\textwidth]{%
            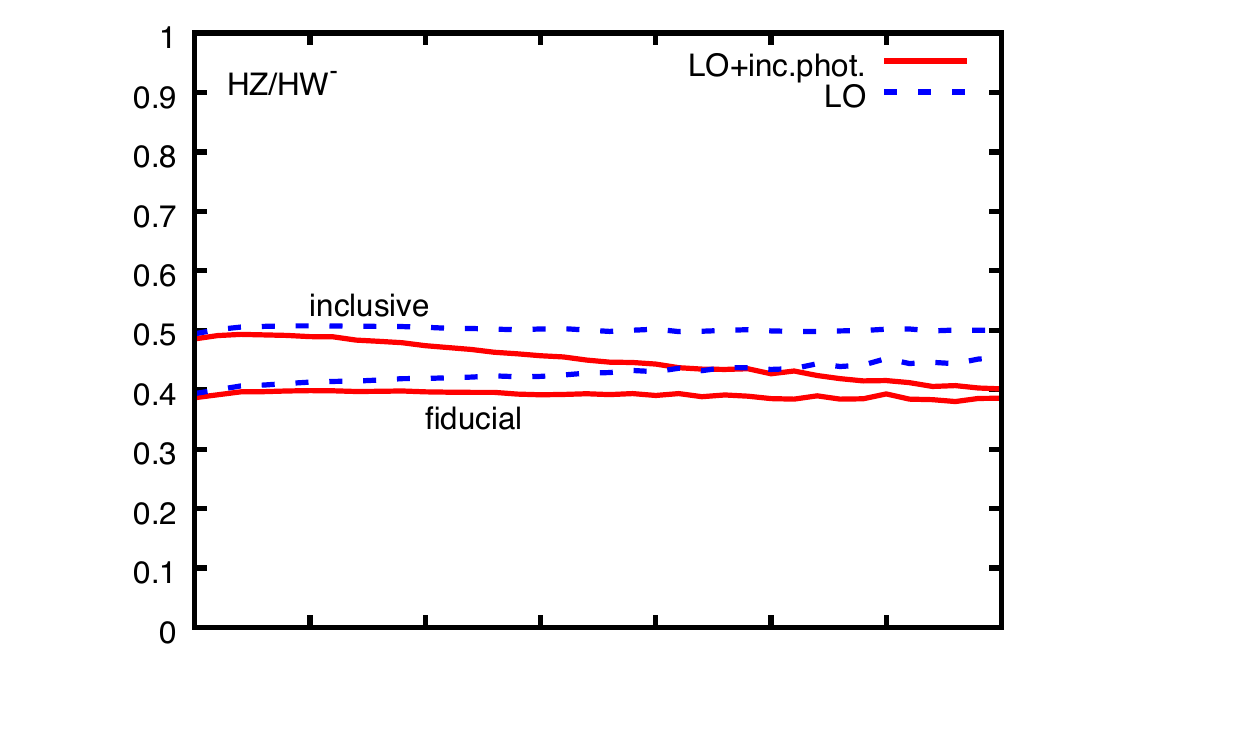}\\[0em]
          \includegraphics[viewport=30 30 300 100,width=.45\textwidth]{%
            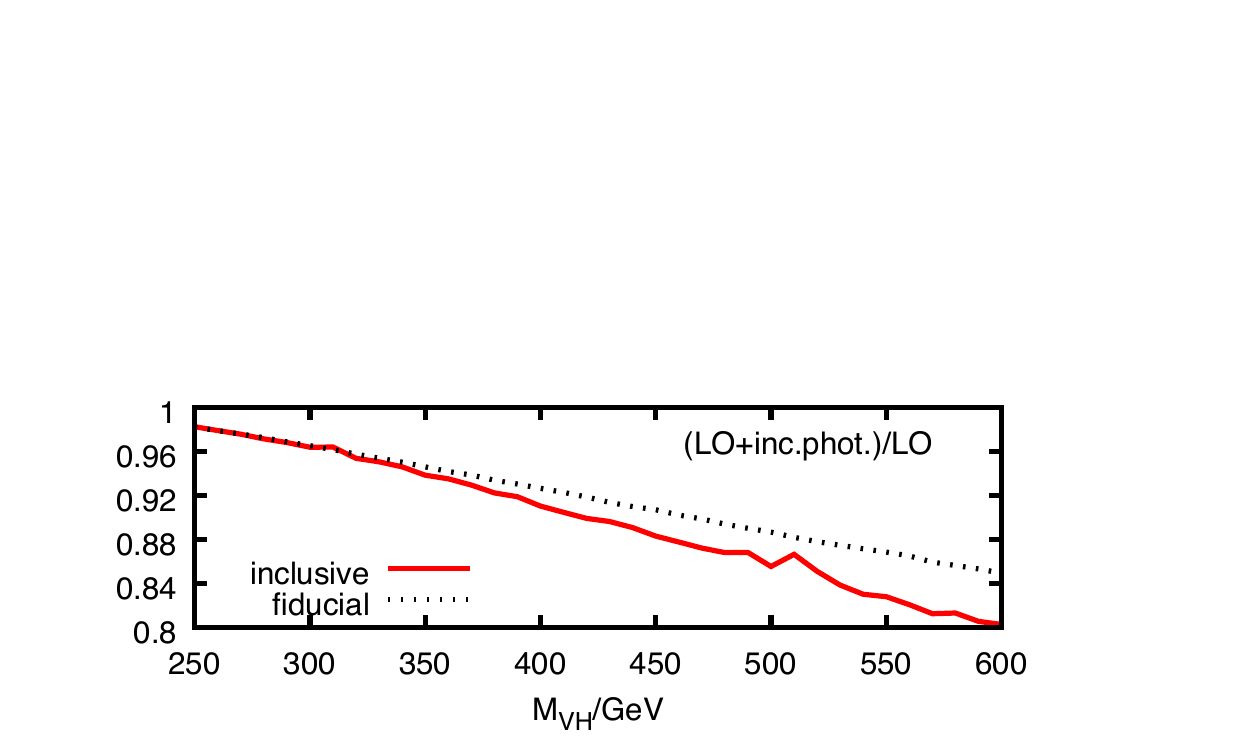}
        \end{tabular}}\\[9em]
      (a) & (b)
    \end{tabular}
    \parbox{.9\textwidth}{
      \caption[]{\label{fig:zw_photon}\sloppy Same as
        \fig{fig:zw_nloew}, but for photon-induced corrections
        $\sigma_\gamma$, i.e.\ $\gamma q\to q\vh$, and using
        \texttt{LUXqed\_plus\_PDF4LHC15} \pdf{}s with
        $\alpha_s(M_Z)=0.118$\,\cite{Manohar:2017eqh,Butterworth:2015oua}.}}
  \end{center}
\end{figure}

A particularly subtle electro-weak contribution is due to the
photon-induced process $\gamma q\to q\vh$, referred to as
$\sigma_\gamma$ in what follows.  Despite the fact that $\sigma_\gamma$
amounts to at most about 7\% to the inclusive $\vh$ production cross
section, its effect on the $\mvh$ distribution of the $\zh/\wh$ ratio
reaches the 20\% level at $\mvh=600$\,GeV, as illustrated in
\fig{fig:zw_photon}.\footnote{Recall again that we only consider
  leptonic $Z$ decays; for $Z\to \nu\bar\nu$, the effect of
  $\sigma_\gamma$ on the \dy-ratio is even larger. \fig{fig:zw_photon}
  has been evaluated using the \pdf{}s described in
  \citere{Manohar:2017eqh}.}  In \citere{deFlorian:2016spz}, an
$\order{100\%}$ uncertainty on $\sigma_\gamma$ was estimated due to its
strong dependence on the available \pdf{} sets, implying a percent-level
uncertainty on the total inclusive $\vh$ production cross section.  Due
to recent theoretical progress in the determination of the photon
\pdf{}s\,\cite{Manohar:2017eqh}, this source of uncertainty on $\vh$
production has been significantly reduced to a level which allows us to
neglect it in our analysis\,\cite{Bertone:2017bme}. A variation of the
electro-weak factorization scale by a factor of two around the central
value of $M_\text{V}+\mhiggs$ changes the electro-weak correction factor
(including the photon-induced corrections) by less than 4\% and would
thus be invisible in \figs{fig:zw_nloew} and \ref{fig:zw_photon}.

\begin{figure}
  \begin{center}
    \begin{tabular}{c}
      \includegraphics[height=.3\textheight]{%
                  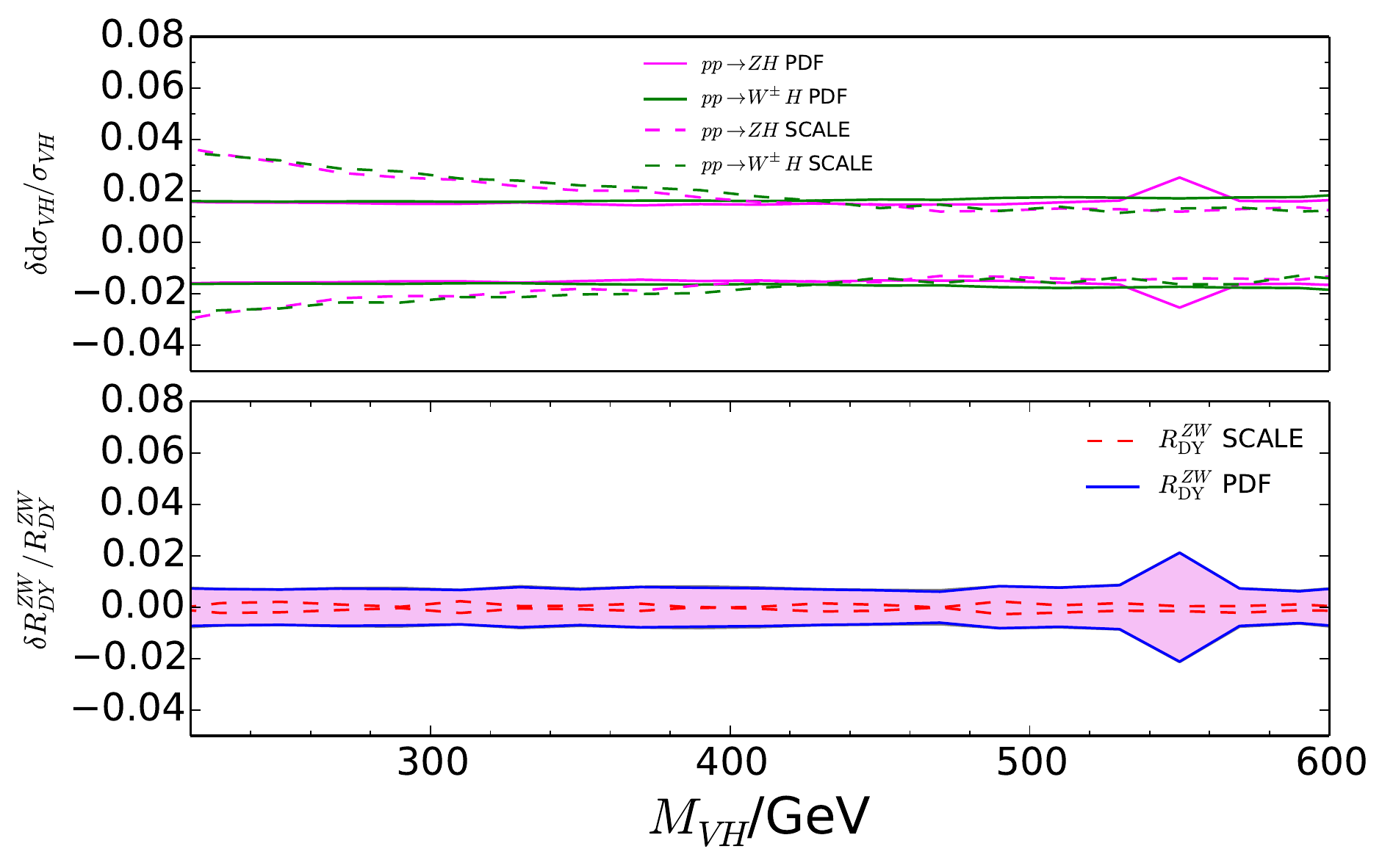}
    \end{tabular}
    \parbox{.9\textwidth}{
      \caption[]{\label{fig:PDFs} \pdf\ uncertainty from the Monte-Carlo
        replicas (using the \texttt{PDF4LHC15\_nlo\_mc} \pdf{} set with
        $\alpha_s(M_Z)=0.118$\,\cite{Butterworth:2015oua}) and
        renormalization/factorization scale uncertainty ($\muF=\muR$
        varied by a factor of two around $\mvh$), evaluated assuming
        full correlation between $\wh$ and $\zh$, and using
        \texttt{MC@NLO} events with one emission added from the parton
        shower. This treatment is formally equivalent to an
        \nlo\ calculation (see main text). }}
  \end{center}
\end{figure}

Let us next consider the uncertainties induced on $\RZWDY{}{}$ by the
\pdf{}s. While they amount to 2-4\% on the cross sections themselves,
they largely cancel in $\RZWDY{}{}$ when assuming that they are fully
correlated between these two processes as demonstrated in
\fig{fig:PDFs}. The uncertainties in this case have been calculated
using \texttt{MadGraph5\_aMC@NLO}~\cite{Alwall:2011uj, Alwall:2014hca}
\abbrev{MC@NLO} events with one emission added through the
\texttt{HERWIG 7} parton shower~\cite{Bellm:2015jjp, Bellm:2017bvx}.
The single parton-shower emission re-introduces the \nlo\ terms
subtracted during the construction of the \abbrev{MC@NLO} events, and
hence this treatment is formally equivalent to an \nlo\ calculation. The
plot also includes the renormalization/factorization scale uncertainty,
obtained by varying these scales by a factor of two around the central
scale, where the latter is defined as half the sum of the transverse
masses of all final state particles (including partons). We assume that
these uncertainties are fully correlated between the $\zh$ and the $\wh$
process, which is justified from the identical form of the \dy-like
\qcd\ corrections for these two processess.  The size of the scale
variation on the ratio corroborates the observations from above about
the uncertainties due to uncalculated higher-order \qcd\ corrections.
The \pdf\ uncertainties for the set
\texttt{PDF4LHC15\_nlo\_mc}~\cite{Butterworth:2015oua} were calculated
using the associated Monte-Carlo replicas. In our analysis below, we
combine these uncertainties in quadrature.

%- {{{ fig:zw_nloqcd

%
%

%- }}}
%- {{{ fig:zw_nloew

%
%

%- }}}
%- {{{ fig:zw_photon
%
%

%- }}}
%- {{{ fig:PDFs:
%
%
%- }}}

%- }}}
%- {{{ subsection{Estimate of the experimental uncertainty}

\subsection{Estimate of the experimental uncertainty}

%- {{{ intro:

In this section, we will provide a rough estimate of
the uncertainty on the double ratio by combining the theoretical
uncertainty on $\RZWDY{}{}$ with the experimental one on $\RZW{}{}$
through \begin{equation}
  \begin{split}
\left(\frac{ \delta \RRZW{}{} } { \RRZW{}{} } \right)^2 = \left(\frac{
  \delta \RZWDY{}{} } { \RZWDY{}{} } \right)_\text{th}^2 + \left(\frac{
  \delta \RZW{}{} } { \RZW{}{} } \right)_\text{exp}^2 \,,
\label{eq:doubleratioerror}
  \end{split}
\end{equation}
where the subscripts indicate that the first term is obtained through a
theoretical calculation and the second through an experimental
measurement. The quadratic sum of theoretical and experimental
uncertainties is justified by the low level of correlation between the
two. We assume total integrated luminosities for $pp$ collisions at
13\,TeV center-of-mass energy of (a)~${\cal L}=36.1$~fb$^{-1}$,
(b)~${\cal L}=300$~fb$^{-1}$, and (c)~${\cal L}=3000$~fb$^{-1}$,
corresponding to (a)~the \abbrev{ATLAS} luminosity underlying the
analysis of \citere{Aaboud:2017xsd}, (b)~the end of \lhc\ Run\,3, and
(c)~the future high-luminosity \lhc\ run.

%- }}}
%- {{{ subsection{Details of the simulated analysis}

\subsubsection{Details of the simulated analysis}\label{sec:simulation}

We construct a hadron-level analysis, including decays of the vector
bosons and the Higgs boson. The parton-level events for signal and
backgrounds are generated at \nlo\ using \texttt{MadGraph5\_aMC@NLO} for
all samples, except for \ggzh\ which is generated at leading order. For
all samples, we employed the \texttt{PDF4LHC15\_nlo\_mc}
\pdf\ set. Parton showering as well as hadronization and modeling of the
underlying event is performed within the general-purpose Monte-Carlo
event generator \texttt{HERWIG 7}. To take into account the higher-order
corrections on \ggzh, we apply a global $K$-factor of
$K=2$\,\cite{Altenkamp:2012sx,Hasselhuhn:2016rqt}.  Electro-weak
corrections largely cancel in the double ratio $R_R^{\zw}$ and can thus
be neglected in our event simulation. We consider leptonic decays of the
vector bosons, $W^\pm \rightarrow \ell^\pm \nu_\ell$ and $Z\rightarrow
\ell^+ \ell^-$, where $\ell = (e, \mu)$, and Higgs-boson decays to
$b\bar b$ pairs.  As background processes we consider $pp \rightarrow
t\bar{t}$, $pp \rightarrow W^\pm b \bar{b}$, $pp \rightarrow Z b
\bar{b}$ and single top production. In this simplified phenomenological
analysis, we do not consider any backgrounds coming from light jets
which are mis-identified as $b$-jets, nor those coming from
mis-identified leptons.\footnote{These are expected to be sub-dominant
  with respect to the `irreducible' backgrounds, as is indeed the case
  in e.g.~\citere{Aaboud:2017xsd}.} To approximately take into account
the \nnlo\ corrections on the $pp \rightarrow t\bar{t}$ background, we
apply a global $K$-factor of $K=1.2$\,\cite{Czakon:2013goa}.

Jets are reconstructed using the anti-$k_T$ algorithm, implemented in
the \texttt{FastJet} package~\cite{Cacciari:2005hq, Cacciari:2011ma}
with parameter $R=0.4$. The jet transverse momentum is required to be
greater than $20$ GeV for `central jets' ($|\eta|<2.5$) and greater than
$30$ GeV for `forward jets' ($2.5<|\eta|<5$). Selected central jets are
labeled as `$b$-tagged' if a $b$-hadron is found within the jet. A
$b$-tagging efficiency of 70\% is considered, flat over the transverse
momentum of the jets, to reproduce the efficiency of the experimental
$b$-tagging algorithm of \citere{Aaboud:2017xsd}. The leading $b$-jet is
required to have a transverse momentum larger than $45$~GeV. The missing
transverse energy is taken as the negative sum of transverse momenta of
all visible particles. Electrons and muons are subject to isolation
criteria by requiring the scalar sum of the transverse momenta of tracks
in $R=0.2$ around them to be less than 1/10$^\text{th}$ of their
transverse momentum: $\sum_{R < 0.2} p_{T}^\mathrm{tracks} < 0.1 \times
p_{T}^\ell$.

\subsubsection{Analysis strategy}\label{sec:analysis}

The 13~TeV \abbrev{ATLAS} analysis of \citere{Aaboud:2017xsd} considered
three event selections, corresponding to the $Z\rightarrow\nu\bar\nu$,
the $W\rightarrow \ell\nu$, and the $Z\rightarrow \ell\ell$
channels. Here we only consider the latter two and refer to them as 1-
and 2-lepton channel, respectively.  All selections require exactly two
$b$-tagged central jets, used to define the invariant mass $m_{b\bar
  b}$. For the $W\rightarrow \ell\nu$ selections, events with more than
three central and forward jets are discarded.

In the $W \to \ell \nu$ analysis, the neutrino four-momentum is
reconstructed by assuming that its transverse component is equal to the
missing transverse momentum, $p^\nu_T = E_{T}^\mathrm{miss}$, and
solving the quadratic equation $(p^\nu + p^\ell)^2 = M_\text{W}^2$ for
the $z$-component $p^\nu_z$. The two resulting solutions can be used to
construct two possible $W$ four-momenta.\footnote{In the case of a
  negative discriminant in the quadratic equation, the
  $E_{T}^\mathrm{miss}$ vector is rescaled such that the discriminant
  becomes zero. The rescaling factor on the two $E_{T}^\mathrm{miss}$
  vector components is chosen to be the same.} These two $W$
four-momenta are then combined with the $b$-jet candidates'
four-momentum, and the combination with invariant mass closest to the
top mass is selected. This invariant mass, denoted by $m_\text{top}$, is
used to suppress top quark-related backgrounds (see last cut below).

Further details on 1-lepton and 2-lepton channels are as
follows:
\begin{description}
\item{$Z\rightarrow ll$ -channel:}
\begin{itemize}
  \item exactly two same-flavor leptons (for muons: of opposite charge)
    with $p_{T}>7$~GeV and $|\eta|<2.5$, of which at least one has
    $p_{T}>25$ GeV;
  \item lepton invariant mass $81$~GeV $< m_{ll}< 101$~GeV;
   \item $p_{T}^{Z} > 150$~GeV
\end{itemize}
\item{$W\rightarrow l\nu$ -channel:}
\begin{itemize}
  \item exactly 1 lepton with $p_{T}>25$ GeV  and $|\eta|<2.5$;
  \item $p_{T}^{W} > 150$ GeV;
  \item $E_{T}^\mathrm{miss} > 30$ GeV in the electron sub-channel;
  \item $m_{b\bar b} > 75$ GeV or $m_\text{top} \leq 225$ GeV.
\end{itemize}
\end{description}

%- {{{ fig:vh_3000_mhv_RDY_hadron

%
\begin{figure}
  \begin{center}
    \begin{tabular}{c}
      \includegraphics[height=.3\textheight]{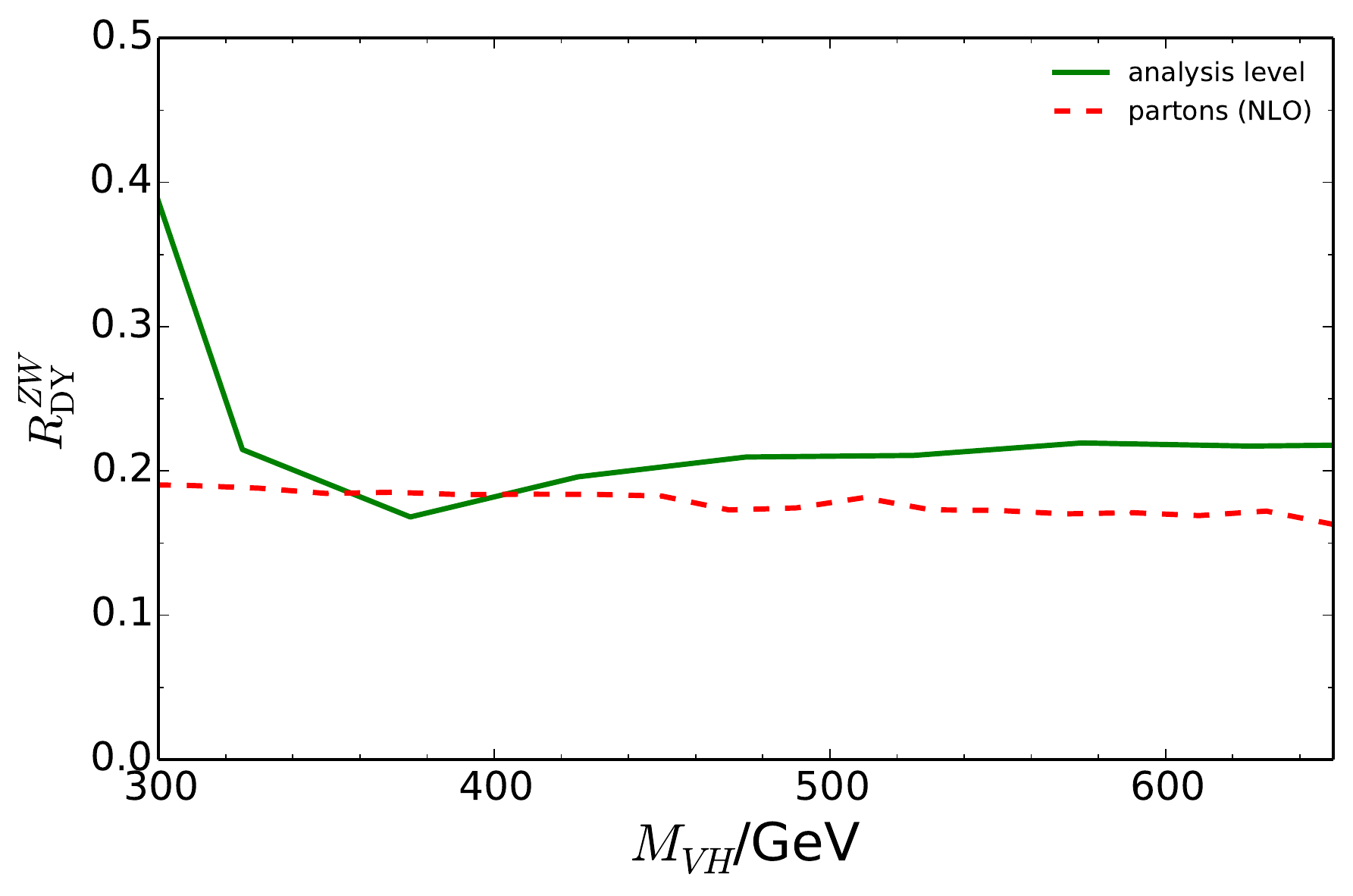}
    \end{tabular}
    \parbox{.9\textwidth}{
      \caption[]{\label{fig:vh_3000_mhv_RDY_hadron}\sloppy Comparison of
        the hadron-level prediction of the ratio of \dy-like
        $\zh$-production to $\wh$-production, $\RZWDY{}{}$, to the
        partonic prediction. (The curves in this plot include the
        branching ratios of the $Z$ and the $W$ boson.) }}
  \end{center}
\end{figure}
%

%- }}}

%In Table~\ref{tab:thcxsecs} we give the initial cross sections, as
%well as those after the analysis is applied. 

The events passing the selection cuts are subject to a ``dijet-mass
analysis'', following closely that of \citere{Aaboud:2017xsd}, where the
\abbrev{BDT}$_{\vh}$ discriminant of the multivariate analysis is
replaced by the invariant mass of the $b$-tagged jets, $m_{b\bar
  b}$. This results in ten signal regions, shown in the second and third
rows of Table~12 in \citere{Aaboud:2017xsd}. In the present analysis, we
have only included signal regions with $p_{T}^{V} > 150$~GeV. We have
further applied the requirement $m_{b\bar b} \in [110, 140]$~GeV which
efficiently selects events containing $H\rightarrow b\bar{b}$.  The
expected number of events predicted by the Monte-Carlo level analysis at
the selection level are similar to those of \citere{Aaboud:2017xsd}.

Figure~\ref{fig:vh_3000_mhv_RDY_hadron} compares the hadron-level
prediction, after analysis cuts, of the ratio $\RZWDY{}{}$ to the
parton-level prediction. The parton-level prediction was constructed
from the truth-level $W$ and Higgs boson momenta, whereas the
hadron-level curve was constructed through the combination of the
reconstructed four-momenta of the $W$ boson and the Higgs boson. For the
$W$ boson, a random choice was made between the two solutions for the
$z$-component of the neutrino momentum.
Figure~\ref{fig:vh_3000_mhv_RDY_hadron} shows that this ratio is only
moderately affected by the analysis and thus can be calculated fairly
reliably within perturbation theory for the inclusive cross section. It
is conceivable that the analysis could be modified appropriately to
preserve more closely the parton-level form of $\RZWDY{}{}$.

%- }}}

\section{Numerical results}\label{sec:numerics}

\subsection{Calculation of experimental uncertainties}\label{sec:uncertainties}

The experimental ratio $\RZW{}{}$ is evaluated from
\begin{equation}\label{eq:Rzweval}
\RZW{}{}= \frac{\mathrm{d} N^{\zh}}{ \mathrm{d} N^{\wh} } = \frac{
  \mathrm{d} N^{\ell \ell} - \mathrm{d} N^{\ell
    \ell}_\mathrm{bkg} }
    {\mathrm{d} N^{\ell} -
  \mathrm{d} N^{\ell}_\mathrm{bkg}}\,,
\end{equation}
where $\mathrm{d} N^{X}$ and $\mathrm{d} N^{X}_\mathrm{bkg}$, with
$X\in\{l,ll\}$, represent the total number of events and the number of
background events per bin, with $Xb\bar b$ final state,
respectively. The uncertainty due to background subtraction will be
included in the estimate of the overall systematic uncertainty. The
uncertainty on \RZW{}{} originating from the size of the total event
samples expected to be collected is given by:
\begin{equation}
\left(\frac{ \delta \RZW{}{} } { \RZW{}{} }\right) _{\mathrm{stat.}}^2 = \left( \frac{ \partial \RZW{}{}  } { \partial(\mathrm{d}  N^{\ell \ell}) }\right)^2 \delta (\mathrm{d} N^{\ell\ell})^2 + \left( \frac{ \partial \RZW{}{} } { \partial (\mathrm{d} N^{\ell}) } \right)^2 \delta (\mathrm{d} N^{\ell})^2 \,.
\end{equation}
If we assume the expected number of events in each bin to be large
enough, then $\mathrm{d} N^{X}$ is Gaussian-distributed with
uncertainty $\delta (\mathrm{d} N^{X} ) = \sqrt{\mathrm{d} N^{X}}$, giving:
\begin{equation}
\left(\frac{ \delta \RZW{}{} } { \RZW{}{} }\right) _{\mathrm{stat.}}^2 =
\frac{ \mathrm{d}N^{\ell\ell} } { (\mathrm{d} N^{\ell\ell} - \mathrm{d}
  N^{\ell \ell}_\mathrm{bkg})^2 } + \frac{ \mathrm{d} N^{\ell} } {
  (\mathrm{d} N^{\ell} - \mathrm{d} N^{\ell}_\mathrm{bkg})^2 } \;.
\end{equation}
We define the systematic uncertainty on $\RZW{}{}$ to include all
uncertainties which contribute to its experimental measurement. A
precise determination of these systematics would require a full-fledged
experimental analysis that would take into account all the correlations
between the different contributing components. For the purpose of this
paper, we content ourselves with an estimate of the uncertainty derived
from the separate $\zh$ and $\wh$ signal strengths of
\eqn{eq:signalstrengths}, presented in the \abbrev{ATLAS} analysis
of~\citere{Aaboud:2017xsd}:
\begin{equation}\label{eq:signalstrengths}
  \begin{split}
\mu_{\zh} &=
1.12^{+0.34}_{-0.33}\mathrm{(stat.)}^{+0.37}_{-0.30}\mathrm{(syst.)}\,,\\ \mu_{\wh}
&=
1.35^{+0.40}_{-0.38}\mathrm{(stat.)}^{+0.55}_{-0.45}\mathrm{(syst.)}\,.
  \end{split}
\end{equation} 
The systematic uncertainty of these results includes all sources of
experimental nature, related to the background and signal Monte-Carlo
simulation and data driven estimates, and to the finite size of the
simulated samples.

We assume that the (symmetrized) systematic uncertainties
$(\delta\mu_{\vh})_\text{syst.}$ can be propagated directly to the
experimental ratio defined by \eqn{eq:Rzweval}, and thus to the double ratio:
\begin{equation}
  \begin{split}
    \left(\frac{\delta R_R^{\zw}}{R_R^{\zw}}\right)^2_\text{syst.} &=
    (\delta\mu_{\zh})_\text{syst.}^2
    +(\delta\mu_{\wh})_\text{syst.}^2-2\,p_{\zw}\,
    (\delta\mu_{\zh})_\text{syst.}(\delta\mu_{\wh})_\text{syst.}\\
    &= 0.112 + 0.250 - 0.335\,p_{\zw}\,.
    \label{eq:syst}
  \end{split}
\end{equation}
where $p_{\zw}$ parameterizes the correlation of the systematic
uncertainties between $\zh$ and $\wh$
production. Table\,\ref{tb:results} shows the results for three
different degrees of correlation: $p_{\zw} = 0$ (no correlation),
$p_{\zw}=1/2$ (50\% correlation), and $p_{\zw}= 1$ (full
correlation).\footnote{An earlier version of these results, based on
  lower statistics of our simulation, has been presented in
  \citere{whzh_lh17}.}

\begin{table}[!htb]
  \begin{center}
    \begin{tabular}{r|c|ccc|ccc}
      && \multicolumn{3}{c|}{stat.\ (${\cal L}/\text{fb}^{-1}$)} &
      \multicolumn{3}{c}{syst.\ ($p_{\zw}$)}\\ & $R_R^{\zw}$ & $36.1$ & $300$ & $3000$ & 0 & 0.5 & 1 \\ \hline 
      all $\mvh$ & 1.49 & $\pm 0.90$ & $\pm 0.31$ & $\pm 0.10$ & $\pm 0.90$ & $\pm 0.66$ & $\pm 0.25$\\ 
      restricted $\mvh$ & 1.55 & $\pm 1.08$ & $\pm 0.38$ & $\pm  0.12$ & $\pm 0.90$ & $\pm 0.66$ & $\pm 0.25$
    \end{tabular}
  \parbox{.9\textwidth}{
    \caption[]{\label{tb:results} Numerical results for the double
      ratio $R^{\zw}_R$ and the associated statistical and systematic
      uncertainties, obtained by mimicking the analysis of
      \citere{Aaboud:2017xsd}. The statistical uncertainty is evaluated
      for three values of the integrated luminosity ($\mathcal{L} =
      36.1$, 300, and 3000\,fb$^{-1}$). The systematic uncertainty is
      shown by assuming the individual systematic uncertainties of $\zh$
      and $\wh$ to be fully uncorrelated, moderately-correlated, and
      fully correlated, respectively ($p_{\zw} = (0, 0.5, 1.0)$). In the
      second line, the $\vh$ invariant mass was restricted to
      $\mvh\in(350,650)$\,GeV. The systematic uncertainties are assumed
      to be unchanged by this restriction.}}
  \end{center}
\end{table}

\subsection{\boldmath Results semi-inclusive in \mvh}\label{sec:inclusiveresults}

Let us first consider integrated quantities before turning to a more
differential analysis below.  From the hadron-level selection described
in Section\,\ref{sec:simulation}, it has been found that the analysis of
\citere{Aaboud:2017xsd} favors events with $\mvh \gtrsim
350$~GeV. Furthermore, we find that, in the present analysis, the
\ggzh\ process contributes substantially up to $\mvh \sim
650$~GeV. Therefore, we also present results where the events are
restricted to $350~\text{GeV} < \mvh < 650~\text{GeV}$. Note that only
the signal regions with $p_{T}^{V} > 150$~GeV are
included.\footnote{Beyond enhancing the \ggzh\ process contribution,
  this also ensures that the 1-lepton and 2-lepton analyses select
  similar phase space regions so as to facilitate cancellation of
  systematic uncertainties.}

Due to the present rudimentary treatment of systematic uncertainties,
these are only considered inclusively, and thus assumed unchanged by
this restriction on the $\mvh$ range. Future experimental analyses,
possessing information on the intricate correlations between systematics
should be able to provide a more differential assessment. We present the
results for the statistical and systematic uncertainties expected at
integrated luminosities of ${\cal L}=36.1 / 300 / 3000$~fb$^{-1}$ in
Table~\ref{tb:results}. From these numbers, one may evaluate the
significance $s$ to which the non-\dy\ component can be observed through
\begin{equation}
  \begin{split}
    s/\sigma &=
    \frac{\sigma^{\zh}_\text{non-\dy}}{\delta\sigma^{\zh}_\text{non-\dy}}
    = \frac{R_R^{\zw}-1}{\delta R_R^{\zw}}=
    \frac{R_R^{\zw}-1}{\sqrt{(\delta R_R^{\zw})^2_\text{stat.}  +
        (\delta R_R^{\zw})^2_\text{syst.}}}\,.
    \label{eq:}
  \end{split}
\end{equation}
For $\mathcal{L}=3000\,\text{fb}^{-1}$, we thus find that the
gluon-initiated component for $\zh$ production as predicted by the
\sm\ gives only a $2\sigma$ effect for the ``restricted \mvh'' sample assuming
full correlation of the systematic errors between $\zh$ and $\wh$
production. In case the systematic uncertainties can be decreased down
to half the current value, the significance increases to
$3.2\sigma$. Considering the fact that New-Physics models typically
enhance the gluon-initiated component, a dedicated experimental analysis
which is tailored to isolate this component and optimized for the
$\zh/\wh$ ratio measurement therefore seems appealing.

Let us take a moment to compare these results to the direct extraction
of the non-\dy\ component from $R_{\dy}^{\zh}$ as sketched at the
beginning of \sct{sec:double_ratio}. In this case, we find a statistical
error of $(\delta R_{\dy}^{\zh})_\text{stat.} = 0.14\,(R_{\dy}^{\zh}-1)$
in the restricted-\mvh\ region, while the systematic error is given by
$(\delta
R^{\zh}_{\dy})_\text{syst.}=R^{\zh}_{\dy}(\delta\mu_{\zh})_\text{syst.}$
if we follow the analogous reasoning as above.  Using our central value
for the double ratio in the restricted-\mvh\ region for $R_{\dy}^{\zh}$, this
leads to a signal significance of $1\sigma$. Assuming that the
systematic uncertainty can be reduced by a factor of two, the
significance for $R_{\dy}^{\zh}\neq 1$ increases to $2\sigma$. Comparing
this to $R_R^{\zw}$, we find that the direct measurement of
$R_{\dy}^{\zh}$ is competitive as long as the correlation between the
systematic $\zh$ and $\wh$ uncertainties is smaller than about 75\%,
i.e.\ roughly the value of $p_{\zw}$ where the correlation term in
\eqn{eq:syst} cancels $(\delta\mu_{\wh})_\text{syst.}$. At this point it
is important to keep in mind that, as argued at the beginning of
\sct{sec:double_ratio}, we also expect significant contributions to the
uncertainty from the theoretical input to $R_{\dy}^{\zh}$, while they
should be negligible for $R_R^{\zh}$. This means that already a
significantly lower $\zh/\wh$ correlation should lead to an improved
extraction of the non-\dy\ contribution by using the double ratio
$R_R^{\zw}$.

\subsection{\boldmath Results differential in \mvh}\label{sec:differentialresults}

\begin{figure}[h]
  \begin{center}
    \begin{tabular}{cc}
      \includegraphics[height=.4\textheight]{%
        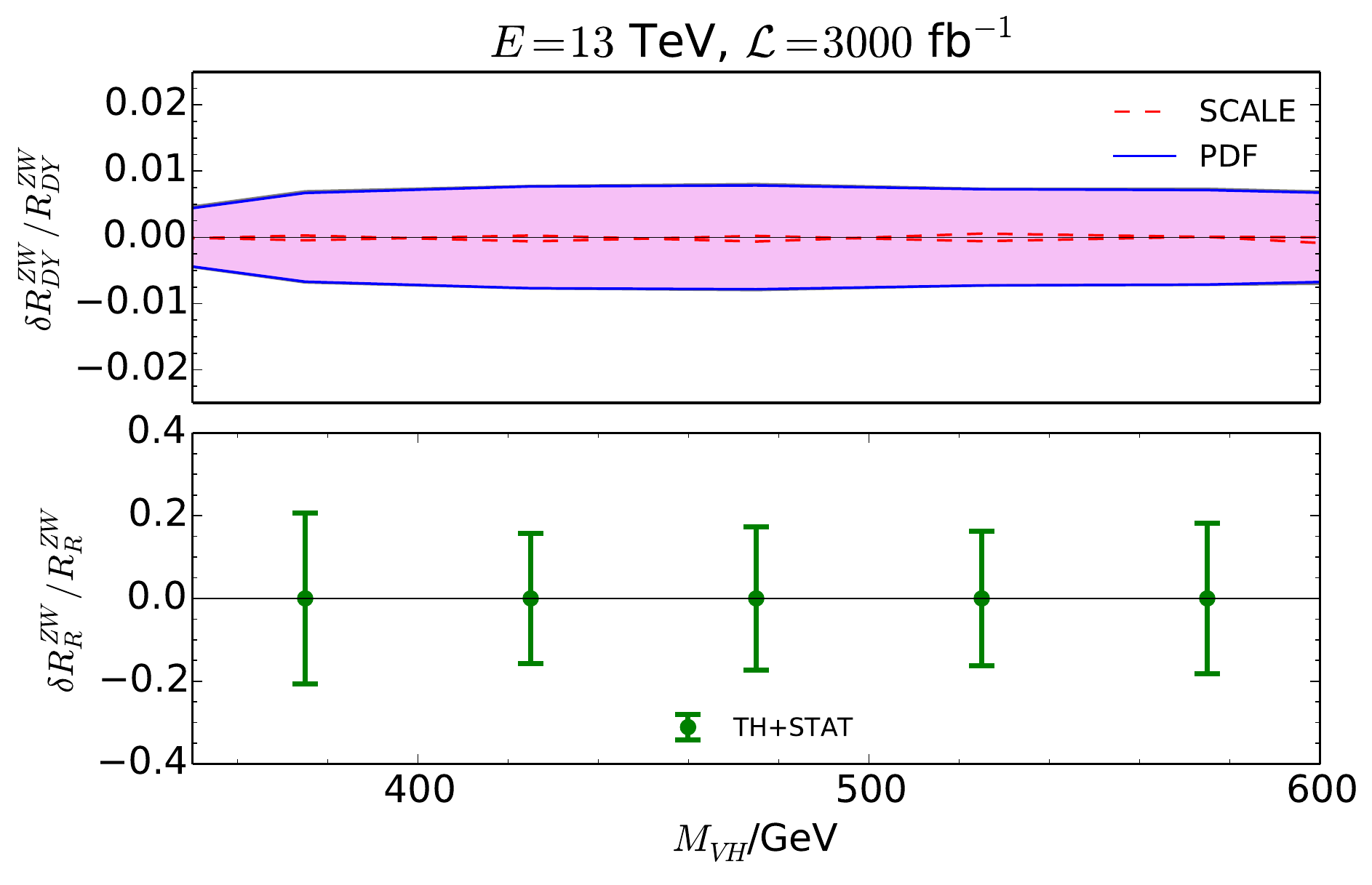} &
    \end{tabular}
    \parbox{.9\textwidth}{
      \caption[]{\label{ggextractratio}\sloppy The upper panel shows the
        ``theoretical'' uncertainty, i.e.\ the first term in
        \eqn{eq:doubleratioerror}, In the lower panel, the green error
        bars show the total relative uncertainty as dictated by
        \eqn{eq:doubleratioerror}. The \sm\ \ggzh\ (with $K=2$) has been
        included in the ``experimental'' uncertainty. The invariant mass
        in the case of the \wh\ channel was constructed through the
        combination of the reconstructed four-momenta of the $W$ boson
        and the Higgs boson. For the $W$ boson, a random choice was made
        between the two solutions for the $z$ component of the neutrino
        momentum.  }}
  \end{center}
\end{figure}

We now turn to the \mvh\ distribution.  Figure\,\ref{ggextractratio}
shows the resulting fractional uncertainties coming from theory or data
statistics as a function of the $\vh$ system invariant mass. The upper
panel shows the ``theoretical'' uncertainty, i.e.\ the first term in
\eqn{eq:doubleratioerror}, obtained by considering the scale and \pdf{}
variations \textit{after} applying the hadron-level analysis. In the
lower panel, the error bars show the total uncertainty as dictated by
\eqn{eq:doubleratioerror}, i.e.\ the combination of the theoretical and
statistical uncertainties for an integrated luminosity of ${\cal
  L}=3000$~fb$^{-1}$, but excluding experimental systematic
uncertainties. We refrain from assessing the latter as their
differential behavior would be challenging to predict at this stage. 
It is evident that the statistical uncertainty originating from the equivalent 
data sample size for an integrated luminosity of ${\cal L}=3000$~fb$^{-1}$, 
$\mathrm{d} N^{X}$, dominates over the theoretical uncertainty.
\begin{figure}[h]
  \begin{center}
    \begin{tabular}{cc}
      \includegraphics[height=.4\textheight]{%
        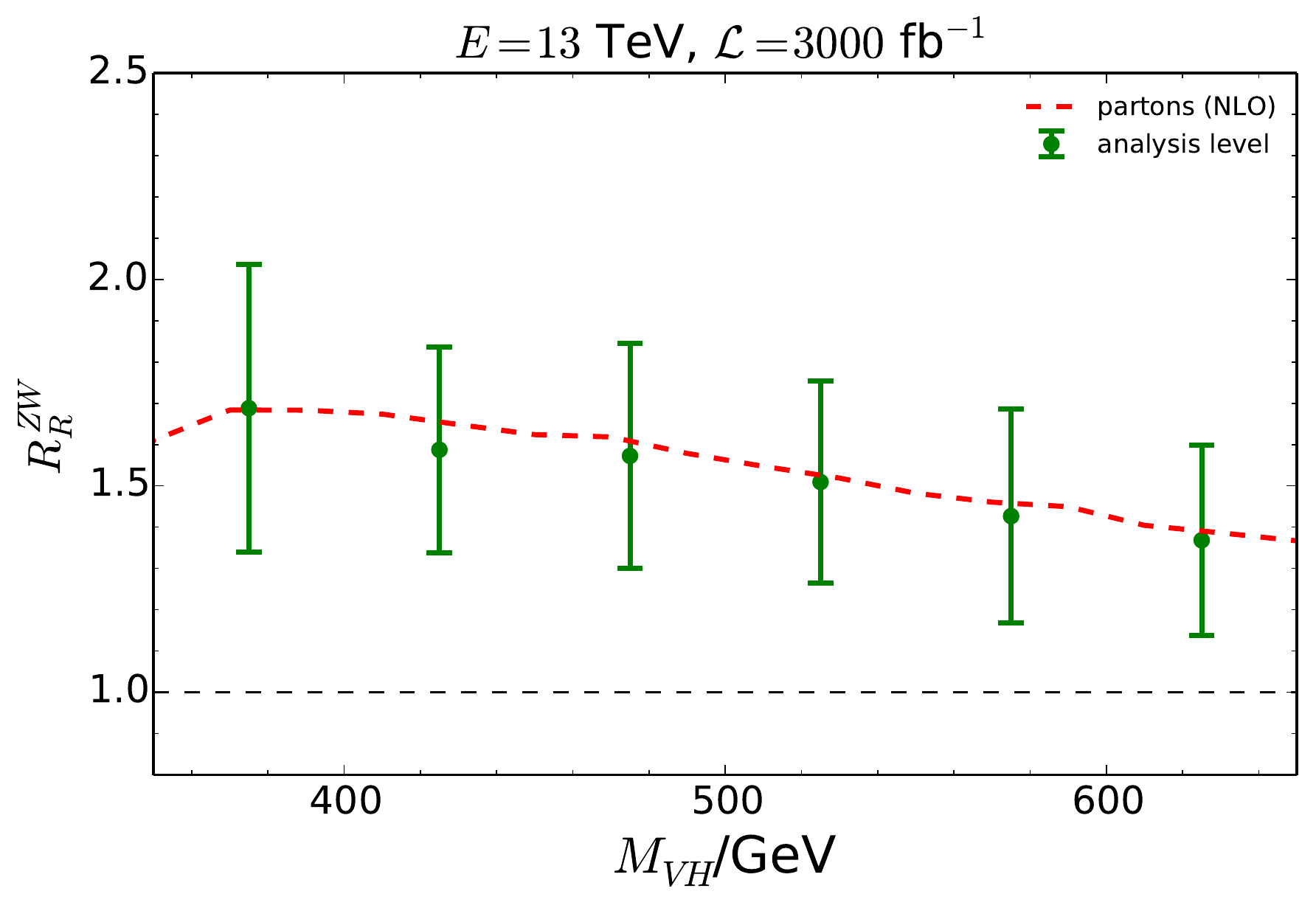} &
    \end{tabular}
    \parbox{.9\textwidth}{
      \caption[]{\label{ggextractratio_absolute}\sloppy The double
      ratio \RRZW{}{} is shown in green error bars, assuming
      \sm\ \ggzh\ (where we have applied a global $K$-factor of
      $K=2$). The size of error bars indicates the total theoretical and
      statistical uncertainty as given
      by~\eqn{eq:doubleratioerror}. The red dashed line shows the
      inclusive double ratio at parton level, including the parton
      shower.}}
  \end{center}
\end{figure}

Figure\,\ref{ggextractratio_absolute} demonstrates how an experimental
measurement would look like, assuming that a \ggzh\ component exists in
the sample at the level of the central \sm\ prediction.\footnote{I.e.,
  we use the \sm\ \ggzh\ cross section at $\muF=\muR=\mvh$, including a
  K-factor of $K=2$.} The double ratio \RRZW{}{} is then given by:
\begin{equation}
\RRZW{}{} = 1 + \frac{ \mathrm{d} N^{\zh}_{gg} } { \mathrm{d}
  N^{\zh}_{\dy}}
\end{equation}
Figure\,\ref{ggextractratio_absolute} also shows the theoretical
parton-level distribution as red dashes (with no cuts applied). The
theoretical prediction and experimental expectation are in good
agreement in this range of $\vh$ invariant mass. Note that the
\abbrev{ATLAS} analysis of \citere{Aaboud:2017xsd} was not constructed
to detect the \ggzh\ component. It is thus conceivable that an analysis
can be devised to increase its contribution to the total $\zh$
production with respect to the parton-level prediction.

\section{Conclusions}

We have investigated New-Physics effects in the gluon-initiated
\higgsstrahlung\ process \ggzh\ and have shown that the $\zh$ invariant
mass distribution provides a particularly sensitive probe for physics
beyond the \sm. While the distribution below the $t\bar t$ threshold,
$\mzh < 2\mtop$, remains rather unperturbed and thus may serve as a
gauge for the experimental data, all New-Physics effects studied here
can be clearly identified and to a large extent even distinguished by
the kinematic region above that threshold. Recall that the low-$\mzh$
region is also under fairly good theoretical control due to existing
higher-order perturbative calculations in the large-$\mtop$
limit\,\cite{Hasselhuhn:2016rqt}. Applying a phenomenological analysis
at the hadronic level in order to estimate the expected theoretical
uncertainty, we find that the \sm\ \ggzh\ component can be established
at the $\sim 3.2\sigma$-level at the \abbrev{HL-LHC} by comparing the
experimental data to the theory prediction for the ratio of \dy-like
$\zh$ production to $\wh$ production in the one- or two-lepton
channels. Adding the zero-lepton channel and optimizing the current
analyses for the \ggzh\ process (or other non-\dy\ processes) would most
likely allow to reveal an $\order{5\sigma}$-level signal.

In order to uniquely establish
New-Physics effects from this method, the theoretical control of the
\ggzh\ component needs to be further increased, for example by including
\sm\ top-mass effects at \nlo. Considering the steady improvement of
theoretical method and existing calculations for very similar processes
(see \citere{Borowka:2016ehy}), it is beyond doubt that
this can be achieved in time for the analysis of \abbrev{HL-LHC} data.

%- }}}
%- {{{ Acknowledgments:

\paragraph{Acknowledgments.} \abbrev{RVH} would like to thank Eric
Laenen and the theory group at Nikhef, where this work was started, for
kind hospitality during the summer of 2017. We would like to thank
Alexander M\"uck for his support in using \hawk{} and Nathan Hartland
for useful discussions on \pdf\ uncertainties. The work of \abbrev{RVH}
and \abbrev{JK} was financially supported by \abbrev{BMBF} under
contract 05H15PACC1. \abbrev{AP} acknowledges support by the
\abbrev{ERC} grant ERC-STG-2015-677323.

%- }}}

%- }}} body:

%- }}}
%- {{{ bibliography:

\def\app#1#2#3{{\it Act.~Phys.~Pol.~}\jref{\bf B #1}{#2}{#3}}
\def\apa#1#2#3{{\it Act.~Phys.~Austr.~}\jref{\bf#1}{#2}{#3}}
\def\annphys#1#2#3{{\it Ann.~Phys.~}\jref{\bf #1}{#2}{#3}}
\def\cmp#1#2#3{{\it Comm.~Math.~Phys.~}\jref{\bf #1}{#2}{#3}}
\def\cpc#1#2#3{{\it Comp.~Phys.~Commun.~}\jref{\bf #1}{#2}{#3}}
\def\epjc#1#2#3{{\it Eur.\ Phys.\ J.\ }\jref{\bf C #1}{#2}{#3}}
\def\fortp#1#2#3{{\it Fortschr.~Phys.~}\jref{\bf#1}{#2}{#3}}
\def\ijmpc#1#2#3{{\it Int.~J.~Mod.~Phys.~}\jref{\bf C #1}{#2}{#3}}
\def\ijmpa#1#2#3{{\it Int.~J.~Mod.~Phys.~}\jref{\bf A #1}{#2}{#3}}
\def\jcp#1#2#3{{\it J.~Comp.~Phys.~}\jref{\bf #1}{#2}{#3}}
\def\jetp#1#2#3{{\it JETP~Lett.~}\jref{\bf #1}{#2}{#3}}
\def\jphysg#1#2#3{{\small\it J.~Phys.~G~}\jref{\bf #1}{#2}{#3}}
\def\jhep#1#2#3{{\small\it JHEP~}\jref{\bf #1}{#2}{#3}}
\def\mpla#1#2#3{{\it Mod.~Phys.~Lett.~}\jref{\bf A #1}{#2}{#3}}
\def\nima#1#2#3{{\it Nucl.~Inst.~Meth.~}\jref{\bf A #1}{#2}{#3}}
\def\npb#1#2#3{{\it Nucl.~Phys.~}\jref{\bf B #1}{#2}{#3}}
\def\nca#1#2#3{{\it Nuovo~Cim.~}\jref{\bf #1A}{#2}{#3}}
\def\plb#1#2#3{{\it Phys.~Lett.~}\jref{\bf B #1}{#2}{#3}}
\def\prc#1#2#3{{\it Phys.~Reports }\jref{\bf #1}{#2}{#3}}
\def\prd#1#2#3{{\it Phys.~Rev.~}\jref{\bf D #1}{#2}{#3}}
\def\pR#1#2#3{{\it Phys.~Rev.~}\jref{\bf #1}{#2}{#3}}
\def\prl#1#2#3{{\it Phys.~Rev.~Lett.~}\jref{\bf #1}{#2}{#3}}
\def\pr#1#2#3{{\it Phys.~Reports }\jref{\bf #1}{#2}{#3}}
\def\ptp#1#2#3{{\it Prog.~Theor.~Phys.~}\jref{\bf #1}{#2}{#3}}
\def\ppnp#1#2#3{{\it Prog.~Part.~Nucl.~Phys.~}\jref{\bf #1}{#2}{#3}}
\def\rmp#1#2#3{{\it Rev.~Mod.~Phys.~}\jref{\bf #1}{#2}{#3}}
\def\sovnp#1#2#3{{\it Sov.~J.~Nucl.~Phys.~}\jref{\bf #1}{#2}{#3}}
\def\sovus#1#2#3{{\it Sov.~Phys.~Usp.~}\jref{\bf #1}{#2}{#3}}
\def\tmf#1#2#3{{\it Teor.~Mat.~Fiz.~}\jref{\bf #1}{#2}{#3}}
\def\tmp#1#2#3{{\it Theor.~Math.~Phys.~}\jref{\bf #1}{#2}{#3}}
\def\yadfiz#1#2#3{{\it Yad.~Fiz.~}\jref{\bf #1}{#2}{#3}}
\def\zpc#1#2#3{{\it Z.~Phys.~}\jref{\bf C #1}{#2}{#3}}
\def\ibid#1#2#3{{ibid.~}\jref{\bf #1}{#2}{#3}}
\def\otherjournal#1#2#3#4{{\it #1}\jref{\bf #2}{#3}{#4}}
\newcommand{\jref}[3]{{\bf #1}, #3 (#2)}
\newcommand{\hepph}[1]{\href{http://arXiv.org/abs/hep-ph/#1}{{\tt hep-ph/#1}}}
\newcommand{\hepth}[1]{\href{http://arXiv.org/abs/hep-th/#1}{{\tt hep-th/#1}}}
\newcommand{\heplat}[1]{\href{http://arXiv.org/abs/hep-lat/#1}{{\tt hep-lat/#1}}}
\newcommand{\mathph}[1]{\href{http://arXiv.org/abs/math-ph/#1}{{\tt math-ph/#1}}}
\newcommand{\arxiv}[2]{\href{http://arXiv.org/abs/#1}{{\tt arXiv:#1}}}
\newcommand{\bibentry}[4]{#1, {\it #2}, #3\ifthenelse{\equal{#4}{}}{}{,}#4.}

%- }}}

\end{document}